\documentclass[aps,prd,amsmath,amssymb,preprintnumbers,preprint,nofootinbib,a4paper]{revtex4-1}
\pdfoutput=1
\usepackage{amsthm}
\usepackage{graphicx}
\usepackage{color}
 \usepackage{colortbl}
	\newcommand{\ccg}{\cellcolor[gray]{0.8}}
	\newcommand{\ccw}{\cellcolor{white}}
	\newcommand{\rcg}{\rowcolor[gray]{0.8}}
\newcommand{\beq}{\begin{eqnarray}}
\newcommand{\eeq}{\end{eqnarray}}

\newcommand{\bmp}{\noindent\begin{minipage}{16cm}}
\newcommand{\emp}{\end{minipage}\vskip 7mm} 

\newcommand{\wt}{\widetilde}

\usepackage{dcolumn}
\usepackage{bm}
\usepackage{bbm}
\usepackage{subfigure}
\usepackage{pxfonts}
\usepackage{epstopdf}
\usepackage{floatpag}

\theoremstyle{definition}

\theoremstyle{plain}

\usepackage{epsfig}

\usepackage[ margin=5pt, font=normalsize,labelfont=bf,justification=raggedright]{caption}
\usepackage{youngtab}
\usepackage{slashed}

\definecolor{rossoCP3}{cmyk}{0,.88,.77,.40}
\def\lsim{\mathrel{\rlap{\lower4pt\hbox{\hskip1pt$\sim$}}
    \raise1pt\hbox{$<$}}}                
\def\gsim{\mathrel{\rlap{\lower4pt\hbox{\hskip1pt$\sim$}}
    \raise1pt\hbox{$>$}}}                

\newcommand{\drawsquare}[2]{\hbox{%
\rule{#2pt}{#1pt}\hskip-#2pt
\rule{#1pt}{#2pt}\hskip-#1pt
\rule[#1pt]{#1pt}{#2pt}}\rule[#1pt]{#2pt}{#2pt}\hskip-#2pt
\rule{#2pt}{#1pt}}

\newcommand{\Yfund}{\raisebox{-.5pt}{\drawsquare{6.5}{0.4}}}

\begin{document}
\includegraphics[width=4.cm]{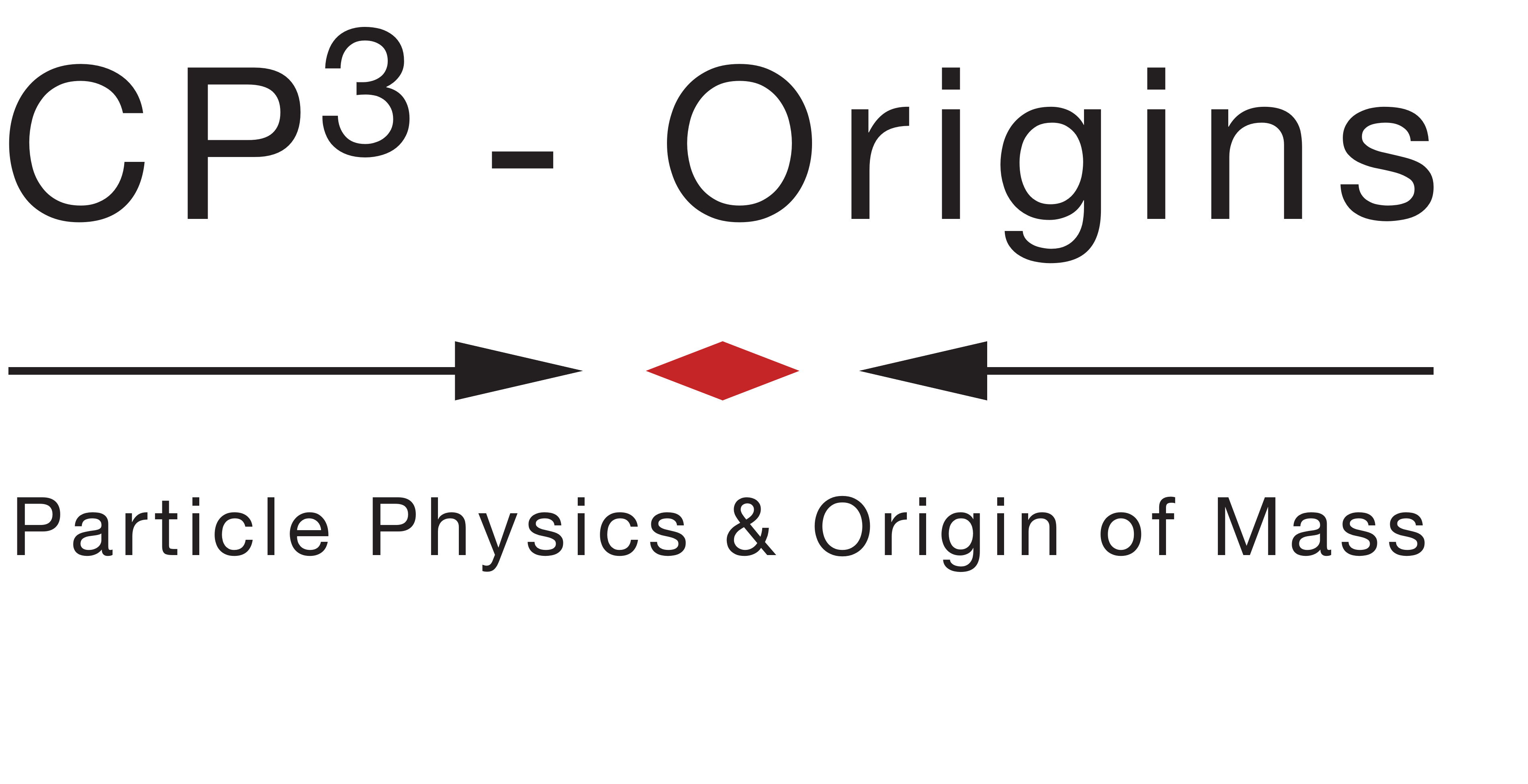}
\title{\Large  \color{rossoCP3} ~~\\Magnetic Fixed Points and Emergent Supersymmetry }
\author{Oleg {\sc Antipin}$^{\color{rossoCP3}{\varheartsuit}}$}\email{antipin@cp3-origins.net} 
\author{Matin {\sc Mojaza}$^{\color{rossoCP3}{\varheartsuit}}$}\email{mojaza@cp3-origins.net} 
\author{Claudio {\sc Pica}$^{\color{rossoCP3}{\varheartsuit}}$}\email{pica@cp3-origins.net} 
\author{Francesco {\sc Sannino}$^{\color{rossoCP3}{\varheartsuit}}$}\email{sannino@cp3-origins.net} 
\affiliation{
{\mbox {$^{\color{rossoCP3}{\varheartsuit}}$
{ \large \rm \color{rossoCP3}CP}$^{\color{rossoCP3}3}${ \large \rm \color{rossoCP3}-Origins}}}  \&  
{{\large D}anish  {\large I}nstitute {\large  f}or  {\large  A}dvanced   {\large  S}tudy}, {\color{rossoCP3}\large \rm DIAS},
 \\
{\mbox{University of Southern Denmark, Campusvej 55, DK-5230 Odense M, Denmark.}}
}

\begin{abstract}
We establish the existence of fixed points for certain gauge theories candidate to be magnetic duals of QCD with one adjoint Weyl fermion. In the perturbative regime of the magnetic theory the existence of a very large number of fixed points is unveiled. We classify them by analyzing their basin of attraction. The existence of several nonsupersymmetric fixed points for the magnetic gauge theory lends further support towards the existence of gauge-gauge duality beyond supersymmetry. We also discover that among these very many fixed points there are supersymmetric ones emerging from a generic nonsupersymmetric renormalization group flow.  We therefore conclude that supersymmetry naturally emerges as a fixed point theory from a nonsupersymmetric Lagrangian without the need for fine-tuning of the bare couplings. Our results suggest that supersymmetry can be viewed as an emergent phenomenon in field theory. In particular there should be no need for fine-tuning the bare couplings when performing Lattice simulations aimed to investigate supersymmetry on the Lattice. 
 \end{abstract}

\maketitle

\section {Introduction}
Gauge theories constitute the building blocks of our present understanding of natural phenomena. The Standard Model (SM) of high energy particle interactions is, in fact, entirely based on a semi-simple gauge group $SU(3)\times SU(2)\times U(1)$. The SM accounts for roughly four percent of the known universe. However, several puzzles remain still unexplained. {}For example, why do we observe, at least, three generations of elementary particles. Besides, the remaining $94 \%$ of the universe, of which $22\%$ dark matter, and $72\%$ dark energy, remains largely unknown. It is therefore natural to expect that new gauge theories, or extensions of the present ones play a fundamental role in explaining the unknown side of the universe. The space of four dimensional gauge theories at our disposal,  without theoretical prejudice, is very large and moreover a large fraction of it is still {\it terra incognita} due to our limited methods to tackle nonperturbative dynamics.  

A fascinating possibility is that different gauge theories display the same physical dynamics. This is possible since the physical quantities are determined by the gauge singlet operators of a generic gauge theory. Therefore it is perfectly legal to imagine to have two different gauge theories leading to the same physics, at least, in certain regimes. Quantum chromodynamics (QCD) itself is a famous example. In fact, at low energy the spectrum can be represented by hadrons interacting via Yukawa interactions and its electric dual description is in terms of the $SU(3)$ gauge theory featuring quarks and gluons. 
 
It is therefore natural to imagine that generic asymptotically free gauge theories have magnetic duals. In fact, in the late nineties, in a series of  ground breaking papers Seiberg \cite{Seiberg:1994bz,Seiberg:1994pq} provided strong support for the existence of a consistent picture of such a duality within a supersymmetric framework. Supersymmetry is, however, quite special and the existence of such a duality does not automatically imply the existence of nonsupersymmetric duals. One of the most relevant  results put forward by Seiberg  has been the identification of the boundary of the conformal window for supersymmetric QCD as a function of the number of flavors and colors. 
Recently  several analytic predictions have been provided for the conformal window of nonsupersymmetric gauge theories using different approaches \cite{Sannino:2004qp,Hong:2004td,Gies:2005as,Braun:2005uj,Braun:2006jd,Dietrich:2006cm,Ryttov:2007sr,Ryttov:2007cx,Sannino:2008ha,Poppitz:2009tw,Sannino:2009za,Braun:2009ns,Antipin:2009wr,Antipin:2009dz,Jarvinen:2009fe,Mojaza:2010cm,Alanen:2010tg,Fukano:2010yv,Pica:2010mt,Pica:2010xq,Frandsen:2010ej,Ryttov:2010iz,Chen:2010er,Ryttov:2010hs,Ryttov:2010jt, Braun:2010qs,Jarvinen:2010ks,Sannino:2009aw,Ryttov:2009yw}. 
We initiated in \cite{Sannino:2009qc}  the exploration of the possible existence of a QCD nonsupersymmetric gauge dual providing a consistent picture of the phase diagram as a function of number of colors and flavors. Arguably the existence of a possible dual of a generic nonsupersymmetric asymptotically free gauge theory able to reproduce its infrared dynamics must match the 't Hooft anomaly conditions \cite{Hooft}. We have exhibited several solutions of these conditions for QCD in \cite{Sannino:2009qc}. An earlier exploration already appeared in \cite{Terning:1997xy}. In \cite{Sannino:2009me} theories with fermions transforming according to higher dimensional representations were analyzed.

In \cite{Mojaza:2011rw}  we have put the idea of nonsupersymmetric gauge duality on a firmer ground by showing that for certain scalarless gauge theories with a spectrum similar to the one of QCD the gauge dual passes a large number of consistency checks. The theory studied  in \cite{Mojaza:2011rw} is QCD with $N_f$ Dirac flavors and one adjoint Weyl fermion. An important feature of this theory is that it possesses the same global symmetry of super QCD despite the fact that squarks are absent. This means that there are four extra anomaly constraints not present in the case of ordinary QCD, moreover we have shown that the potential dual can be constructed for any number of colors greater than two. 

The magnetic dual is a new gauge theory featuring magnetic quarks and a Weyl adjoint fermion, new gauge singlet fermions which can be identified as states composite of the electric variables, as well as scalar states needed to mediate the interactions between the magnetic quarks and the gauge singlet fermions. The new scalars allow for a consistent flavor decoupling which was an important consistency check in the case of supersymmetry. In  \cite{Mojaza:2011rw}  it was also shown that the candidate dual allows to bound the anomalous dimension of the Dirac fermion mass operator to be less than one in the conformal window, and there we also estimated the critical number of flavors below which large distance conformality is lost in the electric variables. 

In these previous studies we implicitly assumed that there exist nontrivial infrared fixed points in the magnetic description matching the electric ones. However, given that we do not start in the ultraviolet with a supersymmetric spectrum nor we impose supersymmetric coupling relations it is legitimate to ask whether these fixed point exist at all in the magnetic description, and if they do what are their universal properties. 

The goal of this work is exactly to establish the existence of these fixed points assuming the most general coupling structure. We will investigate the regime where the computations are trustable, i.e. when the magnetic theory features fixed points accessible in perturbation theory. We discover, for the first time, a large number of calculable fixed points and classify them by analyzing their basin of attraction. An extremely interesting result, according to us, is that among these very many fixed points we discover supersymmetric fixed points to emerge in the infrared along nonsupersymmetric flows. 

Our results have an impact on the (non)supersymmetric gauge-gauge duality and even recent ideas on why the SM can be seen as the magnetic dual description of an electric strongly coupled theory featuring no scalars \cite{Sannino:2011mr}. The consequences of this interpretation of the SM leads to a novel way to tackle the hierarchy problem  and even to shed light on the mystery of the observed number of matter generations  \cite{Sannino:2011mr}.

The phenomenological interest of our studies relies also on the fact that theories similar to the ones investigated here and featuring infrared fixed points have been used to construct sensible extensions of the standard model of particle interactions of technicolor type passing precision data and known as Minimal Walking Technicolor  models \cite{Sannino:2004qp,Dietrich:2006cm}. Duality has also been used to infer relevant insights on the constraints on  these models from precision electroweak tests \cite{Sannino:2010ca,Sannino:2010fh,DiChiara:2010xb}.


\section{Magnetic Setup of QCD with one Adjoint Fermion.}

In \cite{Mojaza:2011rw} we investigated the possible existence of a magnetic dual for the electric theory constituted by a scalarless  $SU(N)$ gauge theory with $N_f$ Dirac fermions and $N$ larger than two, as in QCD, but with an extra Weyl fermion transforming according to the adjoint representation of the gauge group. The quantum global symmetry of the electric theory is therefore:  
\begin{equation}
SU_L(N_f) \times SU_R(N_f) \times U_V(1)  \times U_{AF}(1)\ . 
\end{equation}
At the classical level there is one more $U_A(1)$ symmetry destroyed by quantum
corrections due to the Adler-Bell-Jackiw anomaly.  Of the three independent $U(1)$ symmetries only two survive, a vector like $U_V(1)$ and an axial-like anomaly free (AF) one indicated with $U_{AF}(1)$.  The spectrum of the theory and the global transformations are summarized in table \ref{Electric}.
\begin{table}[t]
\[ \begin{array}{|c| c | c c c c | } \hline
{\rm Fields} &  \left[ SU(N) \right] & SU_L(N_f) &SU_R(N_f) & U_V(1)&U_{AF}(1) \\ \hline \hline
\lambda &{\rm Adj} & 1 &1 &~~0& ~~1 \\
Q &\Yfund &{\Yfund }&1&~~1 & -\frac{N}{N_f} \\
\widetilde{Q} & \overline{\Yfund}&1 &  \overline{\Yfund}& -1 & -\frac{N}{N_f}   \\
G_{\mu}&{\rm Adj}   &1&1  &~~0  & ~~0\\
 \hline \end{array} 
\]
\caption{Field content of the electric theory and field transformation properties. The squared brackets around $SU(N)$ indicate that this is the gauge group.}
\label{Electric}
\end{table}

We then postulated the existence of a magnetic dual featuring the minimal spectrum of composite states and gauge group structure needed to: 

\begin{itemize}

\item{Classify all of the 't Hooft anomaly conditions and show how to match them, for the first time,  for {\it any} number of colors and flavors;}
 
\item{Allow for consistent flavor decoupling both in the electric and in the magnetic theory;} 

\item{Ensure {\it duality involution}. Meaning that if we dualize once more the magnetic theory one recovers the gauge structure of the electric theory.}

\end{itemize}

The proposed  nonsupersymmetric magnetic gauge theory  \cite{Mojaza:2011rw} is summarized in Table \ref{QCDAdual}.
\begin{table}[h]
\[ \begin{array}{|c|c|c c c c|} \hline
{\rm Fields} &\left[ SU(X) \right] & SU_L(N_f) &SU_R(N_f) & U_V(1)& U_{AF}(1) \\ \hline 
\hline 
\lambda_m & {\rm Adj} & 1 & 1 & 0 & 1 \\
 q &\Yfund &\overline{\Yfund }&1&~~\frac{N_f-X}{X} & - \frac{X}{N_f}  \\
\widetilde{q}& \overline{\Yfund}&1 &  {\Yfund}& -\frac{N_f-X}{X}& - \frac{X}{N_f}     \\
 M  & 1 & \Yfund & \overline{\Yfund} & 0 & -\frac{N_f-2X}{N_f} \\
 \hline

  \wt{\phi} & \overline{\Yfund} & 1 & \Yfund & -\frac{N_f-X}{X} & \frac{N_f-X}{N_f}\\
  \phi & \Yfund & \overline{\Yfund} & 1 & \frac{N_f-X}{X} & \frac{N_f-X}{N_f}\\
  G_\mu & \text{Adj} & 1 & 1 & 0 & 0 \\
   \hline \end{array} 
\]
\caption{Field content of the magnetic theory and field transformation properties. The four upper fields are Weyl spinors in the ($1/2,0$) representation of the Lorentz group. The two $\phi$-fields are complex scalars and $G_\mu$ are the gauge bosons.}
\label{QCDAdual}
\end{table}
We have also shown that  one has to have $X=N_f - N$. Differently from the dual of super QCD we did not impose a supersymmetric spectrum or supersymmetric coupling relations. The spectrum above is nonsupersymmetric as it is clear from the fact that there is no complex scalar partner of $M$. Moreover we showed that we could build the gauge singlet states using electric variables which do not contain squarks. Subsequently in \cite{Sannino:2011mr} it was argued that one could also add the complex scalar $H$ (appearing now in Table \ref{QCDAdualm}). This is possible since it does not affect the anomaly conditions and can be built naturally out of the electric fermionic variable as follows \cite{Sannino:2011mr}: 
\begin{equation}
H \sim Q \lambda \lambda \widetilde{Q}  \ .
\end{equation}
It is crucial to be able to construct all these states directly from the electric fermionic variables. This demonstrates that supersymmetry is not a fundamental ingredient in order to construct these states. $H$ in \cite{Sannino:2011mr} plays the phenomenologically relevant role of the SM-like Higgs, elementary in terms of the magnetic variables. Although when adding the new complex scalar field the spectrum of the dual theory looks supersymmetric, the full theory is not since the couplings, in the ultraviolet, are not taken to respect supersymmetric relations. Assuming the duality to exist we were able to make a number of relevant predictions for the nonperturbative gauge dynamics of the strongly coupled QCD gauge theory with one fermion in the adjoint representation. The most relevant ones being: 

\begin{itemize}
\item{The anomalous dimension of the mass of the electric fermions at the lower boundary of the conformal window cannot exceed unity. }

\item{Estimate the size of the conformal window, i.e. the critical number of flavors below which the theory looses large distance conformality.}

\item{When the magnetic theory is used as possible dual of a minimal extension of the SM one can argue that this extension can be re-constructed in terms of a strongly coupled electric theory featuring only fermionic matter, and furthermore argue that the mathematical consistency of the electric dual requires, at least, three generations of ordinary matter \cite{Sannino:2011mr}.}

 \end{itemize}

What was still missing is the actual existence of the fixed points in the magnetic theory. This is what this paper aims to accomplish using perturbation theory. This regime is achieved by choosing  the number of flavors $N_f$, number of colors $X$, and Yukawa couplings where perturbation theory holds.

\begin{table}[t]
\[ \begin{array}{|c|c|c c c c|} \hline
{\rm Fields} &\left[ SU(X) \right] & SU_L(N_f) &SU_R(N_f) & U_V(1)& U_{AF}(1) \\ \hline 
\hline 
\lambda_m & {\rm Adj} & 1 & 1 & 0 & 1 \\
 q &\Yfund &\overline{\Yfund }&1&~~\frac{N_f-X}{X} & - \frac{X}{N_f}  \\
\widetilde{q}& \overline{\Yfund}&1 &  {\Yfund}& -\frac{N_f-X}{X}& - \frac{X}{N_f}     \\
 M  & 1 & \Yfund & \overline{\Yfund} & 0 & -\frac{N_f-2X}{N_f} \\
 \hline

  H & 1 & \Yfund & \overline{\Yfund} & 0 & \frac{2X}{N_f}\\
  \wt{\phi} & \overline{\Yfund} & 1 & \Yfund & -\frac{N_f-X}{X} & \frac{N_f-X}{N_f}\\
  \phi & \Yfund & \overline{\Yfund} & 1 & \frac{N_f-X}{X} & \frac{N_f-X}{N_f}\\
  G_\mu & \text{Adj} & 1 & 1 & 0 & 0 \\
   \hline \end{array} 
\]
\caption{Field content of the magnetic theory with the addition of the Higgs-field, $H$ and the field transformation properties. The four upper fields are Weyl spinors in the ($1/2,0$) representation of the Lorentz group.}
\label{QCDAdualm}
\end{table}

\section{Magnetic potential of the theory and beta functions}

We start with listing the Yukawa operators:\begin{align}\label{LY}
\mathcal{L}_Y = y_\lambda \phi^* \lambda_m q + y_{\widetilde \lambda} \widetilde{\phi}^* \lambda_m \widetilde{q} + y_{\widetilde M} \widetilde{\phi} M q + y_M \phi M \widetilde{q} + y_H \wt{q}\,H q + {\rm h.c.}
\end{align}
$\phi^4$-interactions do not affect our results to the perturbative order we are considering and therefore we will not include them here.

This Yukawa-sector contributes to the two-loop beta function 
of the gauge coupling in the following manner:
\begin{align}
\beta(g) &= \frac{d g}{d \ln \mu} = - \beta_0 \frac{g^3}{(4 \pi)^2} - \beta_1 \frac{g^5}{(4 \pi)^4} - \beta_Y \frac{g^3}{(4 \pi)^4} +  \mathcal{O}(g^7) \label{betag}\\[2mm]
\beta_0 &= \frac{11}{3} C_2(G) - \frac{2}{3} \sum_r T(r) N_f(r)- \frac{1}{6}\sum_s T(s) N_f(s)\\
\beta_1 &= \frac{34}{3} C_2(G)^2- \sum_r \left [ \frac{10}{3} C_2(G) + 2 C_2 (r) \right] T(r) N_f(r) -\sum_s \left [ \frac{1}{3} C_2(G) + 2 C_2 (r) \right] T(s) N_f(s)\\
\beta_Y &=  \frac{1}{d(G)} \sum_r\text{Tr}\left[ C_2(r) Y^j {Y_j}^\dagger\right]\label{betaY}
\end{align}
where $r$ denotes the representation of fermions and $s$ denotes the representation of the \emph{real} scalars. $T(\cdot)$ is the trace normalization of the group generators, $C_2(\cdot)$ is the quadratic Casimir of these and $d(G)$ is the dimension of the gauge group.
We refer to Appendix \ref{app:RG} for a careful derivation of the 
Yukawa contribution $\beta_Y$ to the running of the gauge coupling.
The result is:
\begin{align}
\beta (\alpha_g) = - 2 \alpha_g^2 \left [ \beta_0 + \alpha_g \beta_1+ 
 (\alpha_\lambda + \alpha_{\wt{\lambda}} )  \frac{3X^2-1}{4X} N_f + \left (\frac{\alpha_M + \alpha_{\widetilde{M}}}{2} + \alpha_H \right ) N_f^2 \right],
 \end{align}
 with
 \begin{align}\label{betacoefficients}
 \beta_0 = 3 X - N_f , \quad \beta_1 = 6X^2 - 7N_f X + \frac{3N_f}{X},
 \end{align}
 and where we used the notation
\[ \alpha_i \equiv \frac{\mid y_i \mid^2}{(4 \pi)^2} . \] 

We now consider the running of the Yukawa couplings. 
 The one loop beta function for the Yukawa couplings is given by \cite{Machacek:1983fi,Luo:2002ti}:
\begin{align}
(4 \pi)^2 \beta (Y^j) = \frac{1}{2} \left [ Y^\dagger_2(r) Y^j + Y^j Y_2(r)\right]
&+ 2 Y^k Y^\dagger_j Y^k +  \frac{1}{2}Y^k \text{Tr}\left [ Y^\dagger_k Y^j+Y^\dagger_j Y^k\right]
- 3 g^2 \{ C_2(r), Y^j \}.
\end{align}
Here $Y^j$ is the Yukawa coupling matrix defined by
the particular interaction:
\[
\mathcal{L}_Y \sim Y^{j}_{\alpha \beta} \phi_{j}\psi^\alpha \chi^\beta,
\]
where roman indices contract over the scalar gauge-flavor overall index and the greek indices $\alpha, \beta$ are again gauge-flavor  indices but reserved for the Weyl fermions $\psi$ and $\chi$. 
$Y_2(r)$ is the group invariant:
\[
Y_2(r) \equiv Y^{\dagger}_{j} Y^{j}.
\]
{}For each scalar contraction we multiply, row by column, the Yukawa matrices over the fermion indices. We again report the derivation of the beta function for each Yukawa coupling in
Appendix \ref{app:RG}. To the second order in the couplings the set of beta function equations reads:
\begin{align}\label{RGsystem1}
\beta (\alpha_g) = - 2 \alpha_g^2 &\left [ \beta_0 + \alpha_g \beta_1 + 
 (\alpha_\lambda + \alpha_{\wt{\lambda}} )  \frac{3X^2-1}{4X} N_f + \left(\frac{\alpha_M + \alpha_{\widetilde{M}} }{2} + \alpha_H\right)N_f^2 \right]\\
 \label{RGsystem2}
 \beta(\alpha_{\lambda}) = 2 \alpha_{\lambda} &\left [\frac{3 (X^2-1)}{4X} \alpha_\lambda+ ( \alpha_\lambda + \alpha_{\wt{\lambda}})\frac{ N_f }{4} + N_f \left ( \frac{\alpha_{\wt{M}}+\alpha_H}{2} + \alpha_{M} \right) - 3\alpha_g \frac{3X^2-1}{2X} \right]
 \nonumber\\& 
 -4 N_f\frac{ y_M y_{\widetilde M} y_\lambda y_{\widetilde \lambda}}{(4\pi)^2}
 \\
 \label{RGsystem3}
 \beta(\alpha_{M}) = 2 \alpha_{M} &\left [\frac{3 N_f + X}{2} \alpha_M+ \frac{ X }{2}\alpha_{\wt{M}} + \frac{N_f}{2}\alpha_H +\frac{X^2-1}{2X} \left ( \frac{\alpha_{\wt{\lambda}}}{2} +\alpha_\lambda \right) - 3\alpha_g  \frac{X^2-1}{2X}\right]
 \nonumber\\& 
 -4\frac{X^2-1}{2X} \frac{ y_M y_{\widetilde M} y_\lambda y_{\widetilde \lambda}}{(4\pi)^2}
 \\
 \beta(\alpha_H) = 2 \alpha_H &\left[ \frac{\alpha_M + \alpha_{\wt{M}} + 2\alpha_H}{2}N_f
 + \left (\frac{\alpha_{\lambda} + \alpha_{\wt{\lambda}}}{2}-6\alpha_g\right)C_2(\Yfund) + X \alpha_H \right]\\
 \beta(\alpha_{\wt{\lambda}}) = \beta_{\alpha_{\lambda}}& \left( y_{\lambda}  \leftrightarrow  y_{\widetilde \lambda}  ,  y_{\widetilde M}  \leftrightarrow  y_{M}  \right), \\
 \beta(\alpha_{\widetilde{M}}) = \beta_{\alpha_{M}}&\left(  {y}_{M}  \leftrightarrow  y_{\widetilde M} ,  y_{\widetilde \lambda}  \leftrightarrow  {y}_{\lambda}  \right),
 \label{RGsystem4}
\end{align}
where:
\begin{align}\label{betarelation}
\beta(\alpha_i) \equiv \frac{2 {y_i} }{(4 \pi)^2}\beta(y_i) \ .
\end{align}
 We impose $CP$ invariance and therefore set all the Yukawa phases to zero. We discuss the full set of equations, including the phases, in the Appendix \ref{app:RG}. 
%


In the supersymmetric limit  $\alpha_{\lambda} = \alpha_{\wt{\lambda}} = 2 \alpha_g$,  $\alpha_M = \alpha_{\wt{M}} = \alpha_H$ and the beta function system simplifies to:
\begin{align}\label{SUSYlimit1}
\beta (\alpha_g) 
 &= - 2 \alpha_g^2 
 \left [ \beta_0 + \beta_1^{\prime} \alpha_g + 2N_f^2 \alpha_M \right]\\
 \beta(\alpha_{\lambda}) 
&= - 2 \alpha_{\lambda} \left [ \beta_0 \alpha_g   \right]\\
 \beta(\alpha_{M}) &
 = 2 \alpha_{M} \left [ (2N_f +X) \alpha_M - 4\alpha_gC_2(\Yfund) \right]\\
 \beta(\alpha_H) &= 2 \alpha_H \left[ (2N_f +X) \alpha_H
 - 4\alpha_gC_2(\Yfund)\right],\label{SUSYlimit4}
\end{align}
where $\beta_1^{\prime}\equiv\beta_1+(3X^2-1)N_f/X=6X^2-4N_fX+2N_f/X \hspace{1mm}$  is the well-known two-loop term of pure super QCD, i.e. {\it without} the gauge-singlet chiral superfield transforming as bilinear with respect to the non-abelian global symmetries and, if present, leading to the introduction of a  superpotential term. The theory with this chiral superfield included is Seiberg's magnetic dual.

This result shows, as expected, that supersymmetry
stays unbroken along the renormalization flow.
Also note how the running of the gaugino Yukawa coupling has collapsed
to the one-loop result for the running of the gauge coupling.
These results are in agreement with the known result
from supersymmetry. The all-orders supersymmetric beta functions for Seiberg's magnetic dual are:
\begin{align}
\beta^s (\alpha_g) 
&= - 2 \alpha_g^2 \frac{\left [ \beta_0 + N_f\gamma_0 \right]}{1-2X\alpha_g}\nonumber\\
\beta^s(\alpha_M) 
&=  \alpha_M 
 \left [ \gamma_M + 2\gamma_0 \right],
\end{align}
where the one-loop expressions for the  anomalous dimensions of the chiral superfields read:
\begin{align}
\gamma_0
&=  -4C_2(\Box)\alpha_g+2N_f \alpha_{M} +\dots\nonumber\\
\gamma_M 
&= 2X \alpha_{M} +\dots
\end{align}
by which it is readily seen that the one-loop expansion of $\beta^s(\alpha_g)$
and $\beta^s(\alpha_M)$ are in
agreement with the expressions in equations \eqref{SUSYlimit1}-\eqref{SUSYlimit4}.
\section{Magnetic Fixed Point Analysis }
Just below the critical number of flavors where the theory looses asymptotic freedom ($\beta_0 =0$) 
a Banks-Zaks perturbatively stable infrared fixed point (IRFP) emerges once the Yukawa interactions are set to zero. The well known expression for the value of the gauge coupling at this fixed point is:
\begin{align}
\alpha_g^{*BZ} = - \frac{\beta_0}{\beta_1}\ .
\end{align}
In this phase the theory shows large distance conformality.
We now investigate whether the perturbatively stable IRFP persists when
we turn on the Yukawa interactions.
{}From Eq.\eqref{RGsystem1}, the perturbative fixed point value for the gauge coupling reads:
\begin{align}\label{agfix}
\alpha_g^* = - \frac{4X \beta_0 + (\alpha_\lambda^*+\alpha_{\wt{\lambda}}^*) (3X^2-1) N_f+ 2 X (\alpha_M^* + \alpha_{\wt{M}}^* + 2 \alpha_H^* ) N_f^2 }{4X \beta_1} \ .
\end{align}
Perturbative consistency is ensured by the smallness of the Yukawa couplings at the fixed point.  There is, clearly, a symmetry between the tilded and untilded Yukawa coupling constants at the beta functions level. This symmetry persists at the fixed point effectively reducing  the space of solutions.

The fixed points associated to zeros of $\beta(\alpha_H)$ are:
\begin{align}\label{asfix}
\alpha_H^*=0 \quad \wedge \quad \alpha_H^* = \frac{12  C_2(\Box) \alpha_g^* -C_2(\Box) (\alpha_{\lambda}^*+\alpha_{\wt{\lambda}}^*)-  N_f(\alpha_M^*+\alpha_{\wt{M}}^*)}{2 (N_f + X)}.
\end{align}

A perturbative nontrivial infrared fixed point for  all the couplings is achieved when the number of flavors are such that asymptotic freedom is almost lost. To express this point mathematically we introduce the small parameter $\epsilon$ as follows: 
\[
\beta_0 \geq 0 \Rightarrow N_f \leq 3X, \qquad N_f \equiv 3X (1-\epsilon) \ .
\] 
We report in Table~\ref{solutions0} and  Table~\ref{solutions1} the explicit solutions for the fixed points as a function of $\epsilon$. There are 18 physical solutions for the system \eqref{RGsystem1}-\eqref{RGsystem4}, where $10$ of these correspond to the class $\alpha_H^* =0$ and the remaining $8$ to $\alpha_H^* \neq 0$. The solutions obtained exchanging tilded couplings with untilted ones 
\[
 \alpha_\lambda \leftrightarrow \alpha_{\wt{\lambda}}  \quad \alpha_M \leftrightarrow \alpha_{\wt{M}} \ ,
\]
are indicated in the Tables ~\ref{solutions0} and \ref{solutions1} by an asterisk.
\begin{table}[!hp]
\caption{Fixed point solutions on the critical surface $\alpha_H^* = 0$.
 Solution 
2,3 and 6 are doubled by symmetry property (*).
Note that to linear order in $\epsilon$ solution 5 is the super QCD fixed point.}
\[ \begin{array}{| c |  l | l |} \hline
\text{ \#} &  (*) \, \alpha_\lambda \leftrightarrow \alpha_{\wt{\lambda}} , \quad \alpha_M \leftrightarrow \alpha_{\wt{M}} &  \mathcal{O}(\epsilon)\text{-expansion }
 \\ \hline
1 & \alpha_g^* = \frac{(3X -N_f ) X}{N_f ( 7 X^2-3)-6 X^3} & = \frac{X\epsilon}{5X^2-3} \\
& \alpha_{\lambda, \wt{\lambda}, M, \wt{M}}^* = 0 & \\[2mm] \hline
2^{(*)} &  
\alpha_g^* = \frac{2 ( 3 X-N_f ) ( X (N_f + 3 X)-3)X}{ 2 N_f^2 X ( 7 X^2-3) + 3 N_f (5 - 14 X^2 + X^4)-36 X^3 ( X^2-1)} 
& = \frac{2 X ( 2 X^2-1) \epsilon}{5 - 16 X^2 + 11 X^4} \\
& \alpha_{\lambda}^* = \frac{12 ( 3 X-N_f ) ( 3X^2-1)X}{ 2 N_f^2 X ( 7 X^2-3) + 3 N_f (5 - 14 X^2 + X^4)-36 X^3 ( X^2-1)} 
& =\frac{4 X ( 3 X^2-1) \epsilon}{5 - 16 X^2 + 11 X^4}\\
&  \alpha_{\wt{\lambda}, M ,\wt{M}}^* =0 & \\[2mm] \hline
3^{(*)} & 
\alpha_g^* = \frac{2( 3 X-N_f ) ( 3N_f +X )X}{ 3 N_f^2 (13 X^2-5)-12 X^4 - 2 N_f X (3 + 11 X^2)} 
& = \frac{20X \epsilon}{91 X^2-51} \\
& \alpha_{M}^* = \frac{6 ( 3 X-N_f ) ( X^2-1)X}{ 3 N_f^2 (13 X^2-5)-12 X^4 - 2 N_f X (3 + 11 X^2)} 
& =\frac{6 (  X^2-1) \epsilon}{X(91X^2-51)}\\
&  \alpha_{\lambda,\wt{\lambda} ,\wt{M}}^* =0 & \\[2mm] \hline
4 &
\alpha_g^* = \frac{( 3 X-N_f ) ( 3N_f +2X )X}{2(  N_f^2 ( 9 X^2-3) -6 X^4- N_f X (3 + 2 X^2))} 
& = \frac{11X \epsilon}{2(23 X^2-12)} \\
& \alpha_{M}^* = \alpha_{\wt{M}}^*= \frac{3( 3 X-N_f ) ( X^2-1)}{ 2(  N_f^2 ( 9 X^2-3) -6 X^4- N_f X (3 + 2 X^2))} 
& =\frac{3(  X^2-1) \epsilon}{2X(23X^2-12)}\\
&  \alpha_{\lambda,\wt{\lambda}}^* =0 & \\[2mm] \hline
5 & \alpha_g^* = \frac{( 3 X-N_f ) ( 3X^2+2N_f X -3 )X}{2(  9 X^3 - 9 X^5 + N_f^2 X ( 7 X^2-3) + N_f (3 - 6 X^2 - 9 X^4))} 
& = \frac{X \epsilon}{2(X^2-1)} \\
& \alpha_{\lambda}^* = \alpha_{\wt{\lambda}}^*= \frac{3( 3 X-N_f ) ( 3X^2-1)X}{ 2(  9 X^3 - 9 X^5 + N_f^2 X ( 7 X^2-3) + N_f (3 - 6 X^2 - 9 X^4))} 
& =\frac{X\epsilon}{X^2-1} = 2\alpha_g^* + \mathcal{O}(\epsilon^2)\\
&  \alpha_{M,\wt{M}}^* =0 & \\[2mm] \hline
6^{(*)} &  
\alpha_g^* = \frac{2 ( 3 X-N_f) ( 3 ( N_f^2-1) X -8N_f+ 9 N_f X^2 +3 X^3)X}{ 3 N_f^3 X ( 13 X^2-5)-36 X^4 (X^2-1) + 3 N_f X (5 + 18 X^2 - 31 X^4) + 2 N_f^2 (18 - 59 X^2 + 9 X^4))} 
& = \frac{2 X ( 19 X^2-9) \epsilon}{41 - 141 X^2 + 100 X^4}\\
& \alpha_{\lambda}^* = \frac{12( 3 X-N_f ) (3X^3-2N_f-X+8N_fX^2) }{ 3 N_f^3 X ( 13 X^2-5)-36 X^4 (X^2-1) + 3 N_f X (5 + 18 X^2 - 31 X^4) + 2 N_f^2 (18 - 59 X^2 + 9 X^4))} 
& =\frac{4X ( 27 X^2-7)) \epsilon}{41 - 141 X^2 + 100 X^4}\\
& \alpha_{\wt{M}}^* = \frac{6( 3 X-N_f ) (N_f X-2)(X^2-1) }{ 3 N_f^3 X ( 13 X^2-5)-36 X^4 (X^2-1) + 3 N_f X (5 + 18 X^2 - 31 X^4) + 2 N_f^2 (18 - 59 X^2 + 9 X^4))} 
& =\frac{( 6 X^2-4) \epsilon}{100X^3-41X}\\
&  \alpha_{\wt{\lambda},M}^* =0 & \\[2mm] \hline
7 &
\alpha_g^* = \frac{( 3 X-N_f) (3 ( N_f^2-1) X  -4 N_f  + 6 N_f X^2 + 3 X^3)}{ 2(9 X^3 - 9 X^5 + N_f^3 ( 9 X^2-3) + N_f^2 (X - 15 X^3) + N_f(3 + 6 X^2 - 21 X^4))} 
& = \frac{X ( 16 X^2-5) \epsilon}{2(4X^4-5X^2+1)}\\
& \alpha_{\lambda}^* =\alpha_{\wt{\lambda}}^*= \frac{3( 3 X-N_f ) (5 N_f X^2 -2 N_f - X + 3 X^3) }{9 X^3 - 9 X^5 + N_f^3 ( 9 X^2-3) + N_f^2 (X - 15 X^3) + N_f(3 + 6 X^2 - 21 X^4)} 
& =\frac{X ( 18 X^2-7)) \epsilon}{4X^4 -5X^2+1}\\
& \alpha_M^*=\alpha_{\wt{M}}^* = \frac{3( 3 X-N_f ) (N_f X+3X^2-2)(X^2-1) }{ 2X(9 X^3 - 9 X^5 + N_f^3 ( 9 X^2-3) + N_f^2 (X - 15 X^3) + N_f(3 + 6 X^2 - 21 X^4))} 
& =\frac{( 3 X^2-1) \epsilon}{X(4X^2-1)}\\[2mm] \hline
\end{array}
\]
\label{solutions0}
\end{table}
\begin{table}[!hp]
\vspace{-5mm}
\caption{Fixed point solutions on the critical surface $\alpha_H^* \neq0$.
 Solutions 
2 and 3 are doubled by symmetry property (*).
Note that to linear order in $\epsilon$ solution 6 is the fixed point of Seiberg's magnetic dual.}\[ \begin{array}{| c |  l | l |} \hline
\text{ \#} &  (*) \, \alpha_\lambda \leftrightarrow \alpha_{\wt{\lambda}} , \quad \alpha_M \leftrightarrow \alpha_{\wt{M}} &  \mathcal{O}(\epsilon)\text{-expansion }
 \\ \hline
1 & \alpha_g^* = \frac{(3X -N_f ) (N_f+X)}{4 N_f^2 X - 6 X^3 + N_f ( X^2-3)} 
& = \frac{4X\epsilon}{11X^2-3} \\
&\alpha_H^* =  \frac{(3X -N_f ) (X^2-1)}{X(4 N_f^2 X - 6 X^3 + N_f ( X^2-3))} 
& = \frac{3(X^2-1)\epsilon}{X(11X^2-3)} \\
& \alpha_{\lambda, \wt{\lambda}, M, \wt{M}}^* = 0 & \\[2mm] \hline
2^{(*)} &  
\alpha_g^* = \frac{(3 X - N_f) (6 X ( X^2-1) + N_f ( 7 X^2 + 2 X N_f-5))}{2 X N_f^2 ( 5 X^2 + 4 X N_f-13) -36 X^3 ( X^2-1) -3N_f ( 4 X^2 + 9 X^4-5) } 
& = \frac{X (15 X^2-7) \epsilon}{5 - 26 X^2 + 21 X^4} \\
&\alpha_H^* =  \frac{3 (3 X - N_f) (X^2-1) ( 3 X^2 + 2 X N_f-5)}{X(2 X N_f^2 ( 5 X^2 + 4 X N_f-13) -36 X^3 ( X^2-1) -3N_f ( 4 X^2 + 9 X^4-5) )} 
& = \frac{(9X^2-5)\epsilon}{X(21X^2-5)} \\
& \alpha_{\lambda}^* = \frac{12 X (3 X - N_f) (3 X^2 + 2 X N_f-1)}{ 2 X N_f^2 ( 5 X^2 + 4 X N_f-13) -36 X^3 ( X^2-1) -3N_f ( 4 X^2 + 9 X^4-5) } 
& =\frac{4 X ( 9 X^2-1) \epsilon}{5- 26 X^2 + 21 X^4}\\
&  \alpha_{\wt{\lambda}, M ,\wt{M}}^* =0 & \\[2mm] \hline
3^{(*)} & 
\alpha_g^* = \frac{(3 X - N_f) (2 X^2 + 8 X N_f + 5 N_f^2)}{
20N_f^3X - 12 X^4 + N_f^2(17X^2-15) - 2 N_f X (17X^2+3)} 
& = \frac{71X \epsilon}{193X^2-51} \\
&\alpha_H^* =  \frac{3 (3 X - N_f) (2 X + 5 N_f)( X^2-1) }{X(20N_f^3X - 12 X^4 + N_f^2(17X^2-15) - 2 N_f X (17X^2+3))} 
& = \frac{51(X^2-1)\epsilon}{X(193X^2-51)} \\
& \alpha_{M}^* = \frac{6 ( 3 X-N_f ) ( X^2-1)}{ 20N_f^3X - 12 X^4 + N_f^2(17X^2-15) - 2 N_f X (17X^2+3)} 
& =\frac{6 (  X^2-1) \epsilon}{X(193X^2-51)}\\
&  \alpha_{\lambda,\wt{\lambda} ,\wt{M}}^* =0 & \\[2mm] \hline
4 &
\alpha_g^* = \frac{( 3 X-N_f ) ( 2N_f +X )(N_f+2X)}{2 (4 N_f^3 X - 6 X^4 + N_f^2 (7 X^2-3) - N_f X (3 + 8 X^2))} 
& = \frac{35X \epsilon}{2(47X^2-12))} \\
&\alpha_H^* =  \frac{3 (3 X - N_f) ( N_f+X)( X^2-1) }{X(4 N_f^3 X - 6 X^4 + N_f^2 (7 X^2-3) - N_f X (3 + 8 X^2))} 
& = \frac{12(X^2-1)\epsilon}{X(47X^2-12)} \\
& \alpha_{M}^* = \alpha_{\wt{M}}^*= \frac{3( 3 X-N_f ) ( X^2-1)}{ 2(  4 N_f^3 X - 6 X^4 + N_f^2 (7 X^2-3) - N_f X (3 + 8 X^2))} 
& =\frac{3(  X^2-1) \epsilon}{2X(47X^2-12)}\\
&  \alpha_{\lambda,\wt{\lambda}}^* =0 & \\[2mm] \hline
5 & \alpha_g^* = \frac{(3 X - N_f) (3 X ( X^2-1) + 2 N_f ( 2 X^2 + X N_f-1))}{2 ( 4 N_f X + 3 X^2-3) (N_f^2 X - 3 X^3 - N_f ( X^2+1))} 
& = \frac{X(11X^2-3) \epsilon}{2(5X^2-1)(X^2-1)} \\
&\alpha_H^* =  \frac{3 (3 X - N_f) ( N_fX-1)( X^2-1) }{X( 4 N_f X + 3 X^2-3) (N_f^2 X - 3 X^3 - N_f ( X^2+1)} 
& = \frac{(3X^2-1)\epsilon}{X(5X^2-1)} \\
& \alpha_{\lambda}^* = \alpha_{\wt{\lambda}}^*= \frac{3X( 3 X-N_f ) ( 3X^2+2XN_f-1)}{ ( 4 N_f X + 3 X^2-3) (N_f^2 X - 3 X^3 - N_f ( X^2+1))} 
& =\frac{X(9X^2-1)\epsilon}{(5X^2-1)(X^2-1)} \\
&  \alpha_{M,\wt{M}}^* =0 & \\[2mm] \hline
6 &
\alpha_g^* = \frac{( 3 X-N_f) (2 N_f + X) (  N_f^2 + 2 N_f X + 3 X^2-3)}{ 2(4 N_f^4 X + 9 X^3 - 9 X^5 + N_f^3 ( X^2-3) - N_f^2 X (3 + 23 X^2) + 3 N_f (1 + 4 X^2 - 9 X^4))} 
& = \frac{7X  \epsilon}{2(X^2-1)}\\
&\alpha_H^* =  \frac{3 (3 X - N_f) ( N_fX-N_f^2+1)( X^2-1) }{X(4 N_f^4 X + 9 X^3 - 9 X^5 + N_f^3 ( X^2-3) - N_f^2 X (3 + 23 X^2) + 3 N_f (1 + 4 X^2 - 9 X^4))} 
& = \frac{\epsilon}{X} \\
& \alpha_{\lambda}^* =\alpha_{\wt{\lambda}}^* \frac{3( 3 X-N_f ) (X + 2 N_f) ( 3 X^2 + X N_f-1) }{4 N_f^4 X + 9 X^3 - 9 X^5 + N_f^3 ( X^2-3) - N_f^2 X (3 + 23 X^2) + 3 N_f (1 + 4 X^2 - 9 X^4))} 
& =\frac{7X \epsilon}{X^2-1} = 2\alpha_g^* + \mathcal{O}(\epsilon^2)\\
& \alpha_M^*=\alpha_{\wt{M}}^* = \frac{3( 3 X-N_f ) (3N_f X+3X^2-2)(X^2-1) }{ 2X(4 N_f^4 X + 9 X^3 - 9 X^5 + N_f^3 ( X^2-3) - N_f^2 X (3 + 23 X^2) +3 N_f (1 + 4 X^2 - 9 X^4))} 
& =\frac{ \epsilon}{X} = \alpha_H^* + \mathcal{O}(\epsilon^2)\\ \hline
\end{array}
\]
\label{solutions1}
\end{table}
Solution (5) in Table~\ref{solutions0} corresponds to the perturbative infrared fixed point for super QCD while the last solution (6) in Table \ref{solutions1} corresponds to the perturbative fixed point for Seiberg's magnetic dual and it is the only one with all nonvanishing couplings. Interestingly, when setting to zero $\alpha_H^* =0$ (solution $7$ in Table \ref{solutions0}) we find yet another fixed point corresponding to the infrared stable fixed point of the non-supersymmetric magnetic dual theory proposed in \cite{Mojaza:2011rw} and shown in Table~\ref{QCDAdual}. This lends further support to the proposed non-supersymmetric duality.

\section{Stability analysis}
The stability of the fixed points reported in Tables~\ref{solutions0} and \ref{solutions1} is determined by first evaluating the following matrix
\begin{equation}
\omega_{ij}=\left.\frac{\partial{\beta}_i}{\partial{g}_j}\right|_{g*}
\end{equation} 
at the fixed point of interest denoted in brief by $g^*$.  If it has only real and positive eigenvalues the fixed point is said to be stable. In the Tables \ref{exponents0} and \ref{exponents1}
we present the eigenvalues corresponding to the fixed points of Tables \ref{solutions0} and \ref{solutions1} respectively for $X=4$ and $\epsilon=0.05$. For convenience, we have marked the cells with negative eigenvalues in gray.

\setlength{\arraycolsep}{0.3pt}
\begin{table}
\caption{Eigenvalues corresponding to the fixed points in Table \ref{solutions0} for $X=4$ and $\epsilon=0.05$ case.}%
\[ \begin{array}{|@{}c@{}|c|c|c|c|c|c|} \hline
\text{ \#} &  \nu_1 & \nu_2&\nu_3 & \nu_4&\nu_5 &\nu_6 \\ \hline
\rcg \ccw ~~1~~ & -0.049& -0.049&-0.031&-0.016&-0.016& \ccw 0.0034\\ [2mm ]\hline
2 & 0.20&\ccg -0.048& \ccg -0.045&\ccg-0.015&0.0063&0.0011\\ [2mm ]\hline
3 &\ccg-0.049&\ccg-0.044& 0.035&\ccg-0.029&\ccg-0.015&0.0036\\ [2mm ]\hline
4 &\ccg -0.064& 0.038&0.030&\ccg -0.027&\ccg -0.023&0.0039\\ [2mm ]\hline
5 & 0.30& 0.092& 0.063& \ccg -0.060&\ccg  -0.058& 0.0089\\ [2mm ]\hline
6 & 0.20& \ccg -0.062&\ccg  -0.043&0.031& 0.021& 0.0065\\ [2mm ]\hline
7 & 3.0& 1.8& 1.4& 1.1& \ccg-0.14& 0.033\\ [2mm ]\hline
\end{array}
\]
\label{exponents0}
\end{table}%
\begin{table}
\caption{Eigenvalues corresponding to the fixed points in Table \ref{solutions1} for $X=4$ and $\epsilon=0.05$ case.}%
\[ \begin{array}{|c|c|c|c|c|c|c|} \hline
\text{ \#} &  \nu_1 & \nu_2&\nu_3 & \nu_4&\nu_5 &\nu_6 \\ \hline
~~1~~ & 0.12&\ccg -0.067&\ccg -0.067&\ccg-0.0073&\ccg-0.0073&0.0058\\ [2mm ]\hline
2 & 0.37&0.17& \ccg-0.078& 0.026& 0.012& 0.00038\\ [2mm ]\hline
3 & 0.12&\ccg-0.067&\ccg-0.064&0.015&\ccg-0.0075&0.0058\\ [2mm ]\hline
4 & 0.12&\ccg-0.074&\ccg-0.053&0.015&0.014&0.058\\ [2mm ]\hline
5 & 0.73& 0.26& 0.18& 0.16&\ccg -0.081& 0.018\\ [2mm ]\hline
6 & 21& 5.4& 4.0& 4.0& 0.77& 0.052\\ [2mm ]\hline
\end{array}
\]
\label{exponents1}
\end{table}

There is a rich structure of fixed points classified according to the relevant directions clearly visible from the list of associated eigenvectors reported in the Appendix \ref{Tables}. The dimensions of the critical surfaces are dictated by the number of positive eigenvalues. Once the unstable directions, in the coupling space, are removed the remaining subset of couplings generate a critical $d$-dimensional surface with $d$ given by the number of positive eigenvalues containing the nontrivial fixed points.  {}As an example consider the first fixed point of Table~\ref{exponents0}. Here we have a critical line with an infrared stable fixed point and five unstable directions. The critical {\it surface} in this case is just a line.

\begin{figure}[h]
\includegraphics[width=0.45\textwidth]{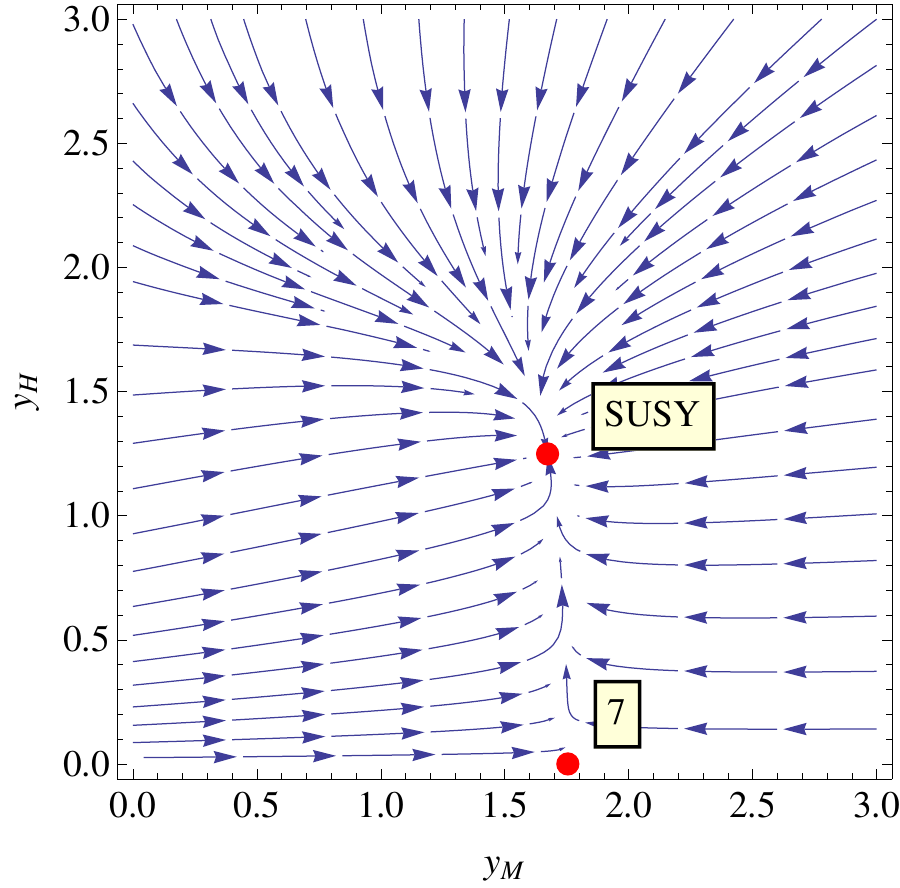} \includegraphics[width=0.45\textwidth]{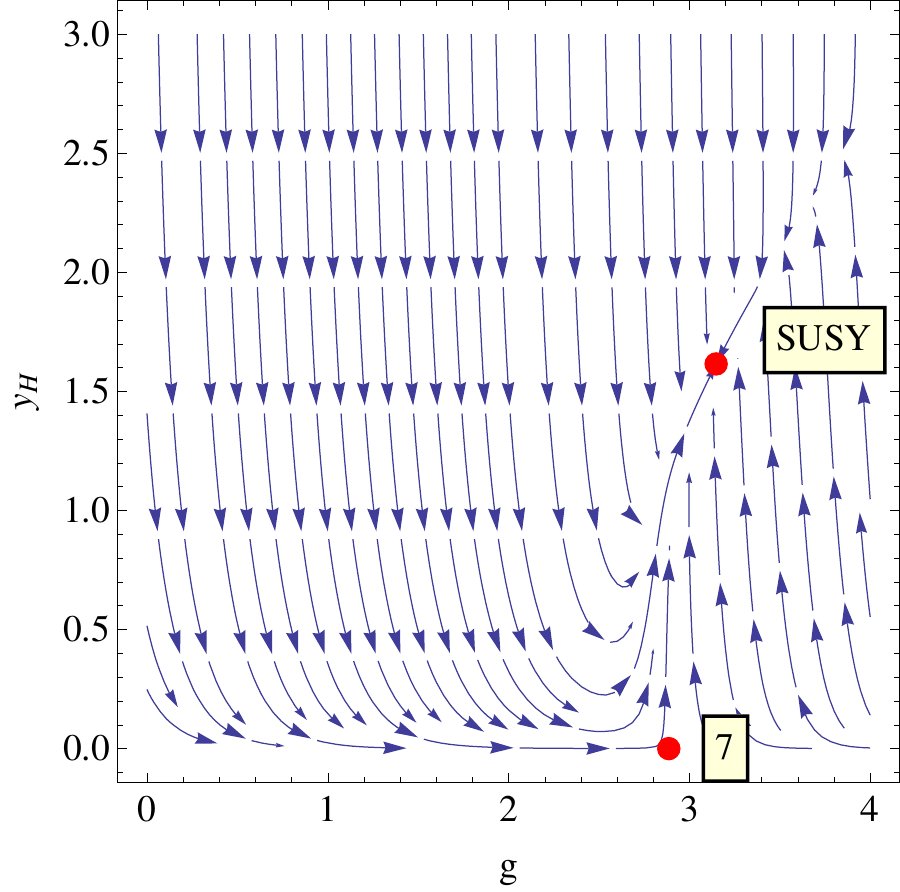} \\ \vspace{2mm}\includegraphics[width=0.45\textwidth]{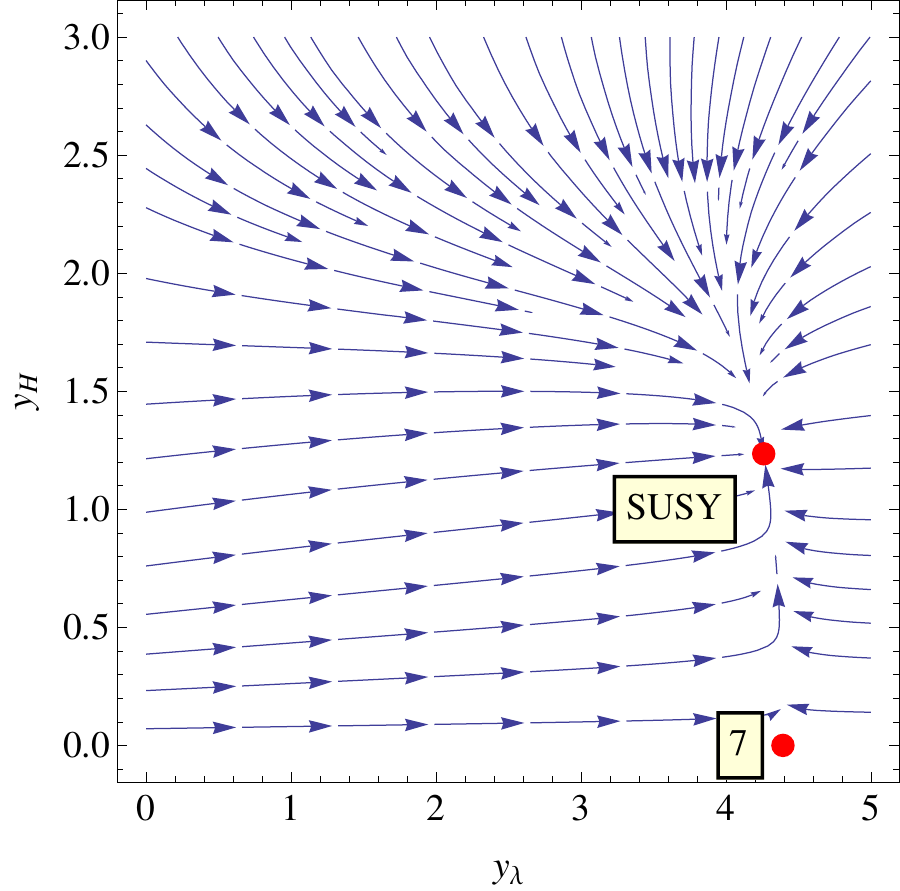} \includegraphics[width=0.45\textwidth]{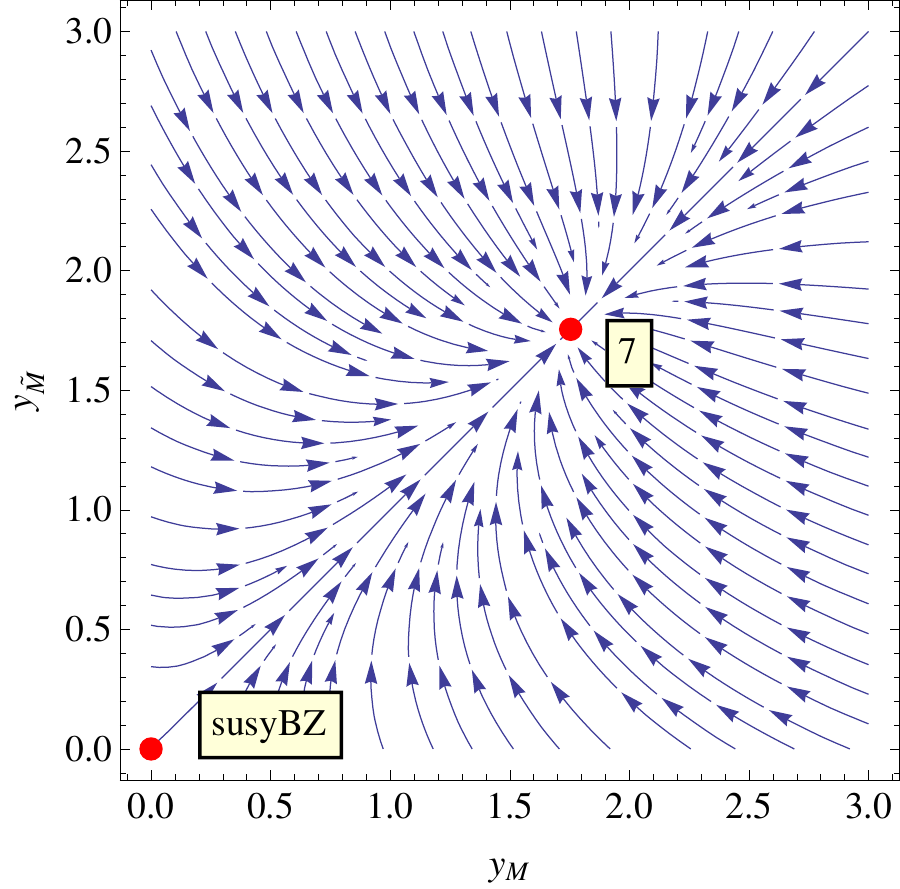}
\caption{ Projections of RG flows for $X=4$ and $\epsilon=0.05$ in the planes (from left to right and top to bottom): ($y_M, y_H$), ($g$, $y_H$), ($y_{\lambda}$,$y_H$) and ($y_M$, $y_{\wt{M}}$). The other couplings are kept fixed at the fixed point value of solution 7 given in Table~\ref{solutions0}. Besides solution 7 clearly indicated in the plot we also plot the projections of the Seiberg's magnetic dual indicated by the SUSY label on the plot, and the super QCD perturbative fixed point indicated by susyBZ. }
\label{FP7}
\end{figure}

{}Another interesting example is constituted by the fixed point solution 7 of Table~\ref{solutions0}. In this case we find a five-dimensional critical surface and one unstable sixth direction parallel to the $y_H$ axis. We plot in Figure \ref{FP7} the projections of the renormalization group flow around this solution  for a subset of two couplings at the time. The other filled small circles in the figure correspond to the projections on the chosen planes of fixed points such as the Seiberg's magnetic dual, indicated by the SUSY label on the plot, or the super QCD Banks-Zaks fixed point labelled by susyBZ.

%
%

{}The fixed point featuring all positive eigenvalues is the one corresponding to Seiberg's magnetic dual.

\section{Physical results and conclusions}

We uncovered the full spectrum of perturbative fixed points associated to an $SU(X)$ nonsupersymmetric gauge theory featuring scalars and fermions, of the type  summarized in Table~\ref{QCDAdualm}. Although the noninteracting field theory has a supersymmetric looking spectrum we did not assume the bare couplings to respect supersymmetry. We then analyzed the fixed points by studying the beta functions to second order in the couplings. To this order, which is renormalization scheme independent, we discovered supersymmetric and nonsupersymmetric fixed points. We further analyzed their stability and discovered that: 

\begin{itemize}
\item  When all bare couplings are nonzero the theory flows to the Seiberg's magnetic dual fixed point.  This occurs on supersymmetric and nonsupersymmetric renormalization flow directions.

\item When the Yukawa couplings $y_M$, $y_{\widetilde M}$ and $y_H$ are all set to zero the theory flows to the super QCD fixed point, as above, independently if the renormalization flow is supersymmetric.  
 
\item When all the Yukawas are set to zero we achieve the perturbative nonsupersymmetric fixed point in the gauge coupling. 

\item When $y_M$ and $y_{\widetilde M}$ are nonzero and $y_H =0$ we discover a new nonsupersymmetric fixed point. This can be identified with the magnetic dual fixed point for nonsupersymmetric gauge theories proposed in \cite{Mojaza:2011rw}. Another relevant fixed point emerges when $y_M = y_{\widetilde M}=0$ and $y_H \neq 0$. 

\item All the other fixed points emerge similarly by setting to zero different Yukawas.   

\end{itemize}

The generic conclusions we draw from our results are: 

\begin{itemize}
\item Supersymmetry naturally emerges as a fixed point theory from a nonsupersymmetric Lagrangian, meaning without fine-tuning the bare couplings. This relevant result demonstrates that supersymmetry can be viewed as an emergent   phenomenon in field theory. Our results, among other things can be used to argue that one does not need to fine-tune the bare couplings when performing Lattice simulations aimed to study supersymmetry on the Lattice \cite{Catterall:2011pd,Catterall:2010nd}.  
 
\item Duality is not a prerogative of supersymmetry given that the theory features many nonsupersymmetric fixed points, depending on the details of the interactions.  Therefore these new fixed points can describe the nonperturbative physics of electric dual gauge theories like the one envisioned in \cite{Mojaza:2011rw}. 
\end{itemize}

The potential physical applications of our results range from (supersymmetric) extensions of the SM to providing new ideas on how to investigate nonperturbative supersymmetric dynamics via first principle lattice simulations.

\acknowledgments

We thank Simon Catterall, Stefano Di Chiara and Marco Nardecchia for enlightening  discussions and comments.

\newpage

\begin{appendix}
\numberwithin{table}{section}
\section{Derivation of the beta functions}\label{app:RG}

We follow the notation of Machacek and Vaughn\cite{Machacek:1983tz,Machacek:1983fi}, 
who derived the general expressions to two-loop order for the running of
the gauge coupling and Yukawa couplings for a general gauge theory
with real scalars and Majorana fermions. 
In Ref. \cite{Luo:2002ti} is showed how to use the expressions
of Machacek and Vaughn for complex scalars and Weyl fermions.
We need only to do a slight rewriting of 
$\mathcal{L}_Y$ given in Eq. \eqref{LY} to follow their notation:
\begin{align}
\mathcal{L}_Y &= 
y_\lambda \phi^* T^a q \lambda_m^a - y_{\widetilde \lambda}\wt{q} T^a \wt{\phi}^* \lambda_m^a + y_M \phi M \wt{q} + y_{\widetilde M} \wt{\phi} M q  + y_H H q \wt{q} + h.c.  \nonumber\\
&=Y^{\alpha,a}_j {\phi^*}^j  q_\alpha \lambda_m^a + 
\widetilde{Y}^j_{\alpha,a} \widetilde{\phi}^*_j \wt{q}^\alpha \lambda_m^a + 
U^j_{\alpha} \phi_j M \wt{q}^\alpha +
\widetilde{U}^\alpha_j \widetilde{\phi}^j {M} {q}_\alpha +
 V^{\alpha}_\beta H q_\alpha \wt{q}^\beta
+ {\rm h.c.},
\end{align}
where $a= 1, \ldots, d(G)$ is the gauge index reserved for the adjoint Majorana fermion with $d(G)$ the dimension of its representation, greek gauge indices $\alpha, \beta, \ldots$ are reserved for the Weyl fermions and roman gauge indices $i,j, \ldots$ are reserved for the complex scalars. Flavor indices have been suppressed.
The Yukawa matrices are defines as follows, once we take flavor indices into account with $l,l'$ denoting the indices of $SU(N_f)_L$ and $r,r'$ indices of the $SU(N_f)_R$ global symmetries:
\begin{subequations}
\begin{align}
Y^{\alpha l, a}_{j l'} &= y_\lambda {\left(T^a\right)_j}^\alpha \delta^{l'}_l, \label{Ymatrix}\\
\widetilde{Y}^{j r'}_{\alpha r, a} &= - y_{\widetilde \lambda} {\left(T^a\right)_\alpha}^j \delta^{r'}_r, \\
U^{j l',r'}_{\alpha r, l} &= y_M  \delta^{j}_\alpha  \delta^{l'}_l \delta^{r'}_r ,  \\
\widetilde{U}^{\alpha l,r}_{j r',l'} &= y_{\widetilde M}   \delta_{j}^\alpha  \delta_{l'}^l \delta_{r'}^r ,\\
V^{\alpha l' , r'}_{\beta r, l} &= y_H \delta^{\alpha}_\beta \delta^{l'}_l \delta^{r'}_r, 
\end{align}
\end{subequations}
where $T^a$ are the generators of the gauge group. 
The minus sign of $y_{\widetilde \lambda}$-term has been chosen to facilitate the connection with supersymmetry, i.e. for SUSY $y_\lambda = y_{\wt{\lambda}} = i \sqrt{2} g$.

We will first derive the Yukawa contribution to the 2-loop beta function, 
given in eq. \eqref{betag}-\eqref{betaY}, in particular:
\begin{align}
\beta_Y &=  \frac{1}{d(G)} \sum_r\text{Tr}\left[ C_2(r) Y^j {Y_j}^\dagger\right] \ ,
\end{align}
where $C_2(r)$ is the quadratic Casimir of the spinor representations $r$ and $d(G)$ is the dimension of the gauge group.
It is instructive to consider the contribution from the Yukawa sector to the two-loop beta function of the gauge coupling diagrammatically. There is only one type of diagram that has non-vanishing contribution, which is:
\begin{center}
\includegraphics[width=0.4\textwidth]{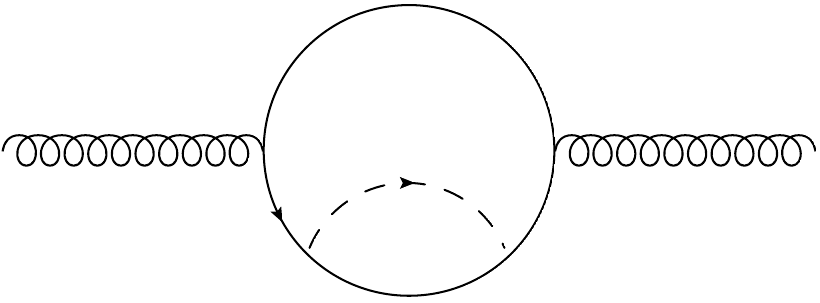}
\end{center}
where the solid line represents spinor fields, and the dashed line represents the scalar(s) coupled via the Yukawa interactions. Note that the scalar-gauge interactions do not contribute.
 
The fermion $M$ is not coupled to the gauge fields. Thus we get:
\begin{align}
\frac{\beta_Y}{(4 \pi)^2 }&= \frac{1}{(4 \pi)^2d(G)}
\text{Tr} \left [ C_2(\lambda)\left \{ Y^j {Y_j}^\dagger + \widetilde{Y}^j {\widetilde{Y}_j}^\dagger \right\}  
 + C_2(q )\left \{ {Y}^j{{Y}_j}^\dagger + \widetilde{U}^j {\widetilde{U}_j}^\dagger  +VV^\dagger \right\} \right. \nonumber\\
 & \qquad \qquad \qquad \qquad \qquad \qquad 
 + \left. C_2(\wt{q})\left \{ \widetilde{Y}^j {\widetilde{Y}_j}^\dagger + U^j {U_j}^\dagger +VV^\dagger  \right\}  
\right] \nonumber\\
&= (\alpha_\lambda + \alpha_{\wt{\lambda}} ) T(\Yfund) N_f \left [ C_2(\Yfund) + C_2(G) \right] + (\alpha_M + \alpha_{\widetilde{M}} +2 \alpha_H) N_f^2 \frac{C_2(\Yfund) X}{d(G)} \nonumber \\
&=  (\alpha_\lambda + \alpha_{\wt{\lambda}} )  \frac{ N_f}{2} \frac{3X^2-1}{2X} + \left (\frac{\alpha_M + \alpha_{\widetilde{M}}}{2} + \alpha_H \right ) N_f^2,
\end{align}
where we used the notation
\[ \alpha_i \equiv \frac{\mid y_i \mid^2}{(4 \pi)^2} . \]
%
  
We now consider the running of the Yukawa couplings. 
The one loop beta function for the Yukawa couplings is given in \cite{Machacek:1983fi,Luo:2002ti}:
\begin{align}
(4 \pi)^2 \beta (Y^j) = \frac{1}{2} \left [ Y^\dagger_2(r) Y^j + Y^j Y_2(r)\right]
&+ 2 Y^k Y^\dagger_j Y^k +  \frac{1}{2}Y^k \text{Tr}\left [ Y^\dagger_k Y^j+Y^\dagger_j Y^k\right]
- 3 g^2 \{ C_2(r), Y^j \}.
\end{align}
where $Y_2(r)$ is the group invariant:
\begin{equation}
Y_2(r) \equiv Y^{\dagger}_k Y^k
\end{equation}
It is again instructive to consider each term in the beta function
in terms of the diagrams they originate from.
The first two terms in the bracket corresponds to the
fermion leg self-energy contribution from the Yukawa sector, 
and the expression tells us that they each contribute with a factor of $\frac{1}{2}$ to the beta function:
\begin{center}
\begin{tabular}{cccc}
\begin{tabular}{c}
 $\frac{1}{2} $ \\ \\ \\ \\ \\
\end{tabular}
&
\includegraphics[width=0.25\textwidth]{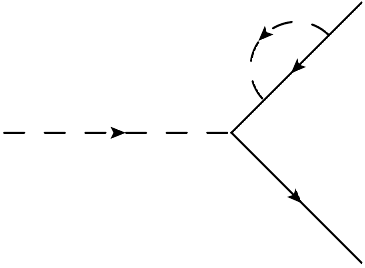}& 
\begin{tabular}{c}
 $+ \frac{1}{2}$ \\ \\ \\ \\ \\
\end{tabular}
\includegraphics[width=0.25\textwidth]{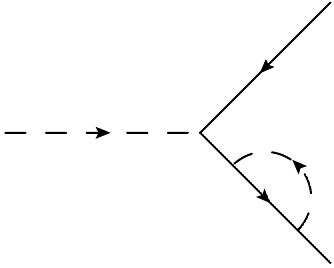}
\end{tabular}
\end{center} 
The next term in the beta function is the vertex correction
from the Yukawa sector:
\begin{center}
\includegraphics[width=0.25\textwidth]{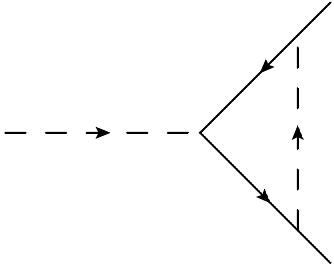}
\end{center}
The fourth term is the scalar leg self energy contribution 
from the Yukawa sector:
\begin{center}
\begin{tabular}{ccc}
\includegraphics[width=0.25\textwidth]{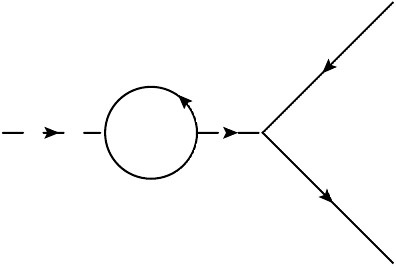} & 
\begin{tabular}{c}
 $+$ \\ \\ \\ \\ 
\end{tabular}
\includegraphics[width=0.25\textwidth]{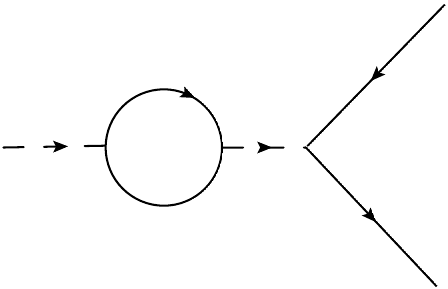} 
\end{tabular}
\end{center} 
Note the different directions of the arrow in the fermion loops.
For Dirac fermions the above contribution comes naturally with a factor of two.
Finally the gauge sector has a non-vanishing contribution only
to the fermion leg self energy, and each term contributes with a factor 
of -3:
\begin{center}
\begin{tabular}{cccc}
\begin{tabular}{c}
 $-3 $ \\ \\ \\ \\ \\ 
\end{tabular}
&
\includegraphics[width=0.25\textwidth]{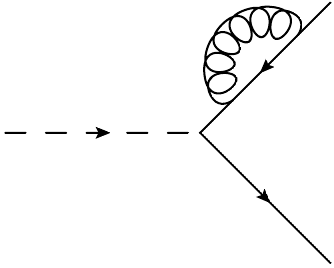}& 
\begin{tabular}{c}
 $-3$ \\ \\ \\ \\ \\ 
\end{tabular}
\includegraphics[width=0.25\textwidth]{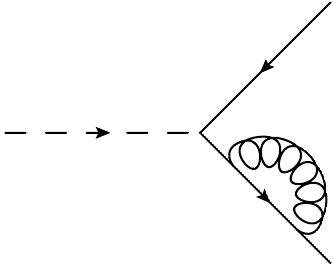}
\end{tabular}
\end{center} 
We thus find the 
one-loop coefficient of the beta function for $Y^j$ to be:
\begin{align}
(4 \pi)^2 \beta (Y^j) = &\frac{1}{2} \left [ \left\{Y^\dagger_2(q)+ \widetilde{U}_2^\dagger(q) + V_2^\dagger(q) \right\} Y^j + Y^j \left\{Y_2(\lambda)+ \widetilde{Y}_2(\widetilde{\lambda})\right\}  \right] + 2 \wt{U}^i U^\dagger_j \wt{Y}^i 
\nonumber\\[2mm]
&+ Y^k \text{Tr}\left [ Y^\dagger_k Y^j\right] + Y^k \text{Tr}\left [ U^k U^\dagger_j\right]
- 3 g^2 \left\{ C_2(q) Y^j + Y^j C_2(\lambda) \right\}.
\end{align}
Inserting the definitions of the Yukawa-matrices, we find:
\begin{align*}
(4 \pi)^2 \beta (Y^j) &= \left[ \frac{1}{2} \left \{C_2(\Yfund)\mid y_\lambda \mid^2 + N_f \mid \tilde{y}_M \mid^2  + N_f \mid y_H \mid^2+ T(\Yfund)N_f(\mid y_\lambda \mid^2 + \mid y_{\widetilde \lambda} \mid^2 ) \right\}
\right. \nonumber \\
&\quad \left. + C_2(\Yfund)\mid y_\lambda \mid^2 + N_f\mid y_M \mid^2 - 3 g^2 \left \{ C_2(\Yfund)+ C_2(G) \right \} \right] Y^j 
-2 N_f y_M^* y_{\widetilde M} y_{\widetilde \lambda} {\left(T^a\right)_\alpha}^j \delta_l^{l'}
%
%
\end{align*}
Note that $ {\left(T^a\right)_\alpha}^j \delta_l^{l'}$ in the last term is the matrix part of $Y^j$, 
which cancels on both sides of the equation once we do the rewriting $\beta(Y^j) =  {\left(T^a\right)_\alpha}^j \delta_l^{l'} \beta(y_\lambda)$.

 The beta function for $\widetilde{Y}^j$ is obtained via:
 \[
 y_\lambda \leftrightarrow \wt{y}_{{\lambda}}  \quad y_M \leftrightarrow \wt{y}_{{M}}
 \]
$\beta(U^j)$ and $\beta(\widetilde{U}^j)$ are related by a similar transformation. 
We compute the former:
\begin{align}
 (4 \pi)^2\beta (U^j) = & 
 \frac{1}{2} \left [ \left\{U^\dagger_2(M)+ \widetilde{U}_2^\dagger(M)\right\} U^j + U^j \left\{U_2(\wt{q})+ \widetilde{Y}_2(\wt{q}) + V_2(\wt{q})\right\}  \right] + 2 \wt{Y}^i Y^\dagger_j \wt{U}^i
 \nonumber\\[2mm]
&+ U^k \text{Tr}\left [ U^\dagger_k U^j\right] + U^k \text{Tr}\left [ Y^\dagger_k Y^j\right]
- 3 g^2 U^j C_2(\wt{q}).
\end{align}
Using the expression for the matrices we deduce:
\begin{align}
\beta (U^j) &=\left[ \frac{1}{2} \left \{X \alpha_M + X \alpha_{\wt{M}} + N_f \alpha_M + C_2(\Yfund) \alpha_{\wt{\lambda}} + N_f \alpha_H \right\} 
 \right.\nonumber \\ &\quad \left. 
 + N_f \alpha_M + C_2(\Yfund) \alpha_\lambda - 3 \alpha_g   C_2(\Yfund)\right] U^j
 - \frac{2 C_2(\Yfund) }{(4\pi)^2} y_{\widetilde \lambda} y_\lambda^* y_{\widetilde M} \delta_\alpha^j \delta_l^{l'} \delta_{r}^{r'},
\end{align}
where we used $\alpha_i = \mid y_i \mid^2/(4 \pi)^2$. As before $ \delta_\alpha^j \delta_l^{l'} \delta_{r}^{r'}$ is the
matrix part of $U^j$, which cancels on both sides of the equation.

The renormalization of the $y_H$ coupling reads:
\begin{align}
(4\pi)^2 \beta(V) = &\left [ \frac{1}{2} \left \{ 
Y^\dagger_2(q)+ \widetilde{U}_2^\dagger(q) + V_2^\dagger(q) \right\} V+ V \left\{U_2(\wt{q})+ \widetilde{Y}_2(\wt{q}) + V_2(\wt{q})\right\}  \right]\nonumber \\
&+ V \text{Tr}\left [ V^\dagger V \right] - 3g^2 \left \{ C_2(q) V + V C_2(\wt{q}) \right \}\nonumber\\[2mm]
\beta(V)=& \left[ \frac{\alpha_M + \alpha_{\wt{M}} + 2\alpha_H}{2}N_f
 + \left (\frac{\alpha_\lambda + \alpha_{\wt{\lambda}}}{2}-6\alpha_g\right)C_2(\Yfund) + X \alpha_H \right] V
\end{align} 

We summarize now the results for beta functions, after having cancelled the matrices:
\begin{subequations}
\begin{align}
&\beta (\alpha_g) = - 2 \alpha_g \left [ \beta_0 + \alpha_g \beta_1 + 
N_f T(\Yfund) \left [ C_2(\Yfund) + C_2(G) \right] (\alpha_\lambda + \alpha_{\wt{\lambda}} ) +  \frac{C_2(\Yfund)X N_f^2}{d(G)} (\alpha_M + \alpha_{\widetilde{M}} +2 \alpha_H) 
 \right]\\
 %
  &\beta(y_\lambda ) =  \mid y_\lambda \mid e^{i\theta_\lambda}
  \left [
 \frac{3}{2} C_2(\Yfund) \alpha_\lambda + \frac{T(\Yfund) N_f}{2} ( \alpha_\lambda + \alpha_{\wt{\lambda}})
 + N_f \left ( \frac{\alpha_{\wt{M}}+\alpha_H}{2} + \alpha_{M} \right)
 - 3\alpha_g  \left \{ C_2(\Yfund)+ C_2(\text{G}) \right \}\right] 
 \nonumber\\& \qquad \quad 
 -2 N_f\frac{\mid y_M \mid \mid y_{\widetilde M}\mid \mid y_{\widetilde \lambda}\mid}{(4\pi)^2}e^{i(\theta_{\wt{\lambda}}+ \Delta \theta_M   )} 
 \\[2mm]
 %
  &\beta( y_M ) = \mid y_M \mid e^{i\theta_M}
  \left [
 \frac{3}{2}N_f \left (\alpha_M + \frac{\alpha_H}{3}\right) + \frac{X}{2} (\alpha_M + \alpha_{\wt{M}})+  C_2(\Yfund) \left ( \frac{\alpha_{\wt{\lambda}}}{2} +\alpha_\lambda \right) - 3\alpha_g C_2(\Yfund)
 \right] \nonumber\\
& \qquad \quad 
 - 2C_2(\Yfund) \frac{\mid y_{\widetilde M}\mid \mid y_\lambda\mid \mid y_{\widetilde \lambda}\mid}{(4\pi)^2}e^{i(\theta_{\wt{M}}+ \Delta \theta_\lambda )} 
 \\[2mm]
 &\beta(\alpha_H) = 2 \alpha_H \left[ \frac{N_f}{2}(\alpha_M + \alpha_{\wt{M}} + 2\alpha_H)
 + C_2(\Yfund) \left (\frac{\alpha_{\lambda} + \alpha_{\wt{\lambda}}}{2}-6\alpha_g\right)+ X \alpha_H \right]\\
 &\beta(\alpha_{\wt{\lambda}}) = \beta_{\alpha_{\lambda}} \left( y_{\lambda}  \leftrightarrow  y_{\widetilde \lambda}  ,  y_{\widetilde M}  \leftrightarrow  y_{M}  \right), \quad
 \beta(\alpha_{\widetilde{M}}) = \beta_{\alpha_{M}}\left(  {y}_{M}  \leftrightarrow  y_{\widetilde M} ,  y_{\widetilde \lambda}  \leftrightarrow  {y}_{\lambda}  \right),
\end{align}
\end{subequations}
where we have made explicit the phases of the Yukawa's couplings:
\begin{align*}
y_i &= \mid y_i \mid e^{i \theta_i} \\
\Delta \theta_M = \theta_{\wt{M}} - \theta_M, \quad & \quad 
\Delta \theta_\lambda = \theta_{\wt{\lambda}} - \theta_\lambda.
\end{align*}
Note also that we define
\[
\beta(\alpha_i ) \equiv \frac{2 \bar {y_i}}{(4\pi)^2}  \beta( y_i ),
\]
where $\bar{y_i}$ is the complex conjugate of $y_i$.

 We have considered the case of real couplings in the main text.

\section{Tables}\label{Tables}
We provide here the eigenvectors associated to the eigenvalues given in Tables \ref{exponents0} and \ref{exponents1}. Shaded cells indicate unstable directions, i.e.  corresponding to negative eigenvalues.
\begin{table}[!hb]
  \begin{center}
\caption{Eigenvectors associated to $\nu_i$ given in Table \ref{exponents0} and corresponding to the fixed points  1,2 and 3 of Table \ref{solutions0}.}%
 \begin{tabular}{ccc}
 $\begin{array}{|c||c|c|c|c|c|c|}\hline
\text{\ref{exponents0}.1}&y_M &y_{\wt{M}}& y_{\lambda}&y_{\wt{\lambda}}&g&y_H\\ \hline \hline
\rcg \nu_1&0 & 0 & 0 & 1.0 & 0 & 0 \\ \hline
 \rcg \nu_2&0 & 0 & 1.0 & 0 & 0 & 0 \\ \hline
\rcg  \nu_3&0 & 0 & 0 & 0 & 0 & 1.0 \\ \hline
\rcg  \nu_4&0 & 1.0 & 0 & 0 & 0 & 0 \\ \hline
\rcg  \nu_5&1.0 & 0 & 0 & 0 & 0 & 0 \\ \hline
 \nu_6&0 & 0 & 0 & 0 & 1.0 & 0 \\ \hline
 \end{array} $
&\hspace{1mm}
$\begin{array}{|c||c|c|c|c|c|c|}\hline
\text{\ref{exponents0}.2}&y_M &y_{\wt{M}}& y_{\lambda}&y_{\wt{\lambda}}&g&y_H\\ \hline \hline
\nu_1&0 & 0 & 0 & -53. & 1.0 & 0 \\ \hline
\rcg \nu_2&0 & 0 & 1.0 & 0 & 0 & 0 \\ \hline
\rcg \nu_3&0 & 0 & 0 & 0 & 0 & 1.0 \\ \hline
\rcg \nu_4&1.0 & 0 & 0 & 0 & 0 & 0 \\ \hline
 \nu_5&0 & 0 & 0 & 1.8 & 1.0 & 0 \\ \hline
 \nu_6&0 & 1.0 & 0 & 0 & 0 & 0 \\ \hline
 \end{array} $
 &\hspace{1mm}
$ \begin{array}{|c||c|c|c|c|c|c|}\hline
\text{\ref{exponents0}.3}&y_M &y_{\wt{M}}& y_{\lambda}&y_{\wt{\lambda}}&g&y_H\\ \hline \hline
\rcg\nu_1&0 & 0 & 1.0 & 0 & 0 & 0 \\ \hline
\rcg \nu_2&0 & 0 & 0 & 1.0 & 0 & 0 \\ \hline
 \nu_3&0 & -46. & 0 & 0 & 1.0 & 0 \\ \hline
 \rcg\nu_4&0 & 0 & 0 & 0 & 0 & 1.0 \\ \hline
 \rcg\nu_5&1.0 & 0 & 0 & 0 & 0 & 0 \\ \hline
 \nu_6&0 & 0.61 & 0 & 0 & 1.0 & 0 \\ \hline
 \end{array}$
\end{tabular}
 \end{center}
\end{table}%
\begin{table}[!hb]
  \begin{center}
\caption{
Eigenvectors associated to $\nu_i$ given in Table \ref{exponents0} and corresponding to the fixed points 4 and 5 of Table \ref{solutions0}.}%
 \begin{tabular}{cc}
$ \begin{array}{|c||c|c|c|c|c|c|}\hline
\text{\ref{exponents0}.4}&y_M &y_{\wt{M}}& y_{\lambda}&y_{\wt{\lambda}}&g&y_H\\ \hline \hline
\rcg\nu_1&0 & 0 & 1.0 & 1.0 & 0 & 0 \\ \hline
 \nu_2&-23. & -23. & 0 & 0 & 1.0 & 0 \\ \hline
 \nu_3&-1.0 & 1.0 & 0 & 0 & 0 & 0 \\ \hline
\rcg \nu_4&0 & 0 & 0 & 0 & 0 & 1.0 \\ \hline
 \rcg\nu_5&0 & 0 & -1.0 & 1.0 & 0 & 0 \\ \hline
 \nu_6&0.58 & 0.58 & 0 & 0 & 1.0 & 0 \\ \hline
 \end{array} $
 &\hspace{2mm}
$ \begin{array}{|c||c|c|c|c|c|c|}\hline
\text{\ref{exponents0}.5}&y_M &y_{\wt{M}}& y_{\lambda}&y_{\wt{\lambda}}&g&y_H\\ \hline \hline
\nu_1&0 & 0 & -22. & -22. & 1.0 & 0 \\ \hline
 \nu_2&0 & 0 & -1.0 & 1.0 & 0 & 0 \\ \hline
 \nu_3&-1.0 & 1.0 & 0 & 0 & 0 & 0 \\ \hline
\rcg\nu_4& 1.0 & 1.0 & 0 & 0 & 0 & 0 \\ \hline
 \rcg\nu_5&0 & 0 & 0 & 0 & 0 & 1.0 \\ \hline
 \nu_6&0 & 0 & 1.5 & 1.5 & 1.0 & 0 \\ \hline
 \end{array}$
 \end{tabular}
 \end{center}
\end{table}%
\begin{table}
  \begin{center}
\caption{Eigenvectors associated to $\nu_i$ given in Table \ref{exponents0} and corresponding to the fixed points 6 and 7 of Table \ref{solutions0}.}%
 \begin{tabular}{cc}
$  \begin{array}{|c||c|c|c|c|c|c|}\hline
\text{\ref{exponents0}.6}&y_M &y_{\wt{M}}& y_{\lambda}&y_{\wt{\lambda}}&g&y_H\\ \hline \hline
\nu_1&-2.2 & 0 & 0 & -49. & 1.0 & 0 \\ \hline
\rcg \nu_2&0 & 0.22 & 1.0 & 0 & 0 & 0 \\ \hline
\rcg \nu_3&0 & 0 & 0 & 0 & 0 & 1.0 \\ \hline
 \nu_4&-40. & 0 & 0 & 13. & 1.0 & 0 \\ \hline
 \nu_5&0 & -0.73 & 1.0 & 0 & 0 & 0 \\ \hline
 \nu_6&0.47 & 0 & 0 & 1.8 & 1.0 & 0 \\ \hline
 \end{array} $
&\hspace{5mm}
$ \begin{array}{|c||c|c|c|c|c|c|}\hline
\text{\ref{exponents0}.7}&y_M &y_{\wt{M}}& y_{\lambda}&y_{\wt{\lambda}}&g&y_H\\ \hline \hline
\nu_1&-0.035 & -0.035 & -3.1 & -3.1 & 1.0 & 0 \\ \hline
 \nu_2&-0.41 & 0.41 & -1.0 & 1.0 & 0 & 0 \\ \hline
 \nu_3&0.40 & -0.40 & -1.0 & 1.0 & 0 & 0 \\ \hline
 \nu_4&-2.0 & -2.0 & 1.7 & 1.7 & 1.0 & 0 \\ \hline
 \rcg\nu_5&0 & 0 & 0 & 0 & 0 & 1.0 \\ \hline
 \nu_6&0.64 & 0.64 & 1.6 & 1.6 & 1.0 & 0\\ \hline
 \end{array}$
\end{tabular}
\end{center}
\end{table}%
\begin{table}
  \begin{center}
\caption{Eigenvectors associated to $\nu_i$ given in Table \ref{exponents1} and corresponding to the fixed points  1 and 2 of Table \ref{solutions1}.}%
 \begin{tabular}{cc}
$\begin{array}{|c||c|c|c|c|c|c|}\hline
\text{\ref{exponents1}.1}&y_M &y_{\wt{M}}& y_{\lambda}&y_{\wt{\lambda}}&g&y_H\\ \hline \hline
\nu_1&0 & 0 & 0 & 0 & -0.052 & 1.0 \\ \hline
\rcg \nu_2&0 & 0 & 0 & 1.0 & 0 & 0 \\ \hline
 \rcg\nu_3&0 & 0 & 1.0 & 0 & 0 & 0 \\ \hline
 \rcg\nu_4&0 & 1.0 & 0 & 0 & 0 & 0 \\ \hline
 \rcg\nu_5&1.0 & 0 & 0 & 0 & 0 & 0 \\ \hline
 \nu_6&0 & 0 & 0 & 0 & 1.1 & 1.0 \\ \hline
 \end{array}  $
&\hspace{5mm}
$\begin{array}{|c||c|c|c|c|c|c|}\hline
\text{\ref{exponents1}.2}&y_M &y_{\wt{M}}& y_{\lambda}&y_{\wt{\lambda}}&g&y_H\\ \hline \hline
\nu_1&0 & 0 & 0 & 5.2 & -0.28 & 1.0 \\ \hline
 \nu_2&0 & 0 & 0 & -1.3 & -0.057 & 1.0 \\ \hline
 \rcg\nu_3&0 & 0 & 1.0 & 0 & 0 & 0 \\ \hline
 \nu_4&0 & 1.0 & 0 & 0 & 0 & 0 \\ \hline
 \nu_5&0 & 0 & 0 & 2.0 & 1.2 & 1.0 \\ \hline
 \nu_6&1.0 & 0 & 0 & 0 & 0 & 0 \\ \hline
 \end{array}  $
\end{tabular}
\end{center}
\end{table}%
\begin{table}
  \begin{center}
\caption{Eigenvectors associated to $\nu_i$ given in Table \ref{exponents1} and corresponding to the fixed points 3 and 4 of Table \ref{solutions1}.}%
 \begin{tabular}{cc}
$\begin{array}{|c||c|c|c|c|c|c|}\hline
\text{\ref{exponents1}.3}&y_M &y_{\wt{M}}& y_{\lambda}&y_{\wt{\lambda}}&g&y_H\\ \hline \hline
\nu_1&0 & 0.15 & 0 & 0 & -0.054 & 1.0 \\ \hline
\rcg \nu_2&0 & 0 & 1.0 & 0 & 0 & 0 \\ \hline
 \rcg\nu_3&0 & 0 & 0 & 1.0 & 0 & 0 \\ \hline
 \nu_4&0 & -6.0 & 0 & 0 & 0.094 & 1.0 \\ \hline
\rcg \nu_5&1.0 & 0 & 0 & 0 & 0 & 0 \\ \hline
 \nu_6&0 & 0.47 & 0 & 0 & 1.2 & 1.0 \\ \hline
 \end{array} $
&\hspace{5mm}
$\begin{array}{|c||c|c|c|c|c|c|}\hline
\text{\ref{exponents1}.4}&y_M &y_{\wt{M}}& y_{\lambda}&y_{\wt{\lambda}}&g&y_H\\ \hline \hline
\nu_1&0.16 & 0.16 & 0 & 0 & -0.057 & 1.0 \\ \hline
 \rcg\nu_2&0 & 0 & 1.0 & 1.0 & 0 & 0 \\ \hline
\rcg \nu_3&0 & 0 & -1.0 & 1.0 & 0 & 0 \\ \hline
 \nu_4&-1.0 & 1.0 & 0 & 0 & 0 & 0 \\ \hline
 \nu_5&-2.9 & -2.9 & 0 & 0 & 0.093 & 1.0 \\ \hline
 \nu_6&0.49 & 0.49 & 0 & 0 & 1.2 & 1.0 \\ \hline
 \end{array}  $
\end{tabular}
\end{center}
\end{table}%
\begin{table}
  \begin{center}
\caption{Eigenvectors associated to $\nu_i$ given in Table \ref{exponents1} and corresponding to the fixed points 5 and 6 of Table \ref{solutions1}.}%
 \begin{tabular}{cc}
$\begin{array}{|c||c|c|c|c|c|c|}\hline
\text{\ref{exponents1}.5}&y_M &y_{\wt{M}}& y_{\lambda}&y_{\wt{\lambda}}&g&y_H\\ \hline \hline
\nu_1&0 & 0 & 3.9 & 3.9 & -0.54 & 1.0 \\ \hline
 \nu_2&0 & 0 & -0.93 & -0.93 & -0.076 & 1.0 \\ \hline
 \nu_3&0 & 0 & -1.0 & 1.0 & 0 & 0 \\ \hline
 \nu_4&-1.0 & 1.0 & 0 & 0 & 0 & 0 \\ \hline
\rcg \nu_5&1.0 & 1.0 & 0 & 0 & 0 & 0 \\ \hline
 \nu_6&0 & 0 & 1.8 & 1.8 & 1.3 & 1.0 \\ \hline
 \end{array}  $
&\hspace{5mm}
$\begin{array}{|c||c|c|c|c|c|c|}\hline
\text{\ref{exponents1}.6}&y_M &y_{\wt{M}}& y_{\lambda}&y_{\wt{\lambda}}&g&y_H\\ \hline \hline
\nu_1&0.41 & 0.41 & 5.1 & 5.1 & -7.4 & 1.0 \\ \hline
\nu_2&-0.44 & 0.44 & -1.0 & 1.0 & 0 & 0 \\ \hline
\nu_3&0.37 & -0.37 & -1.0 & 1.0 & 0 & 0 \\ \hline
\nu_4&1.7 & 1.7 & -3.8 & -3.8 & -1.2 & 1.0 \\ \hline
\nu_5&-0.38 & -0.38 & -0.26 & -0.26 & 0.072 & 1.0 \\ \hline
\nu_6&1.1 & 1.1 & 2.8 & 2.8 & 1.9 & 1.0 \\ \hline
\end{array}  $
\end{tabular}
\end{center}
\end{table}
\end{appendix}


\clearpage
\begin{thebibliography}{30}
\bibitem{Seiberg:1994bz}
 N.~Seiberg,
 Phys.\ Rev.\  D {\bf 49}, 6857 (1994)
 [arXiv:hep-th/9402044].

\bibitem{Seiberg:1994pq}
 N.~Seiberg,
 Nucl.\ Phys.\  B {\bf 435}, 129 (1995)
 [arXiv:hep-th/9411149].
 
 
 
 \bibitem{Sannino:2004qp}
 F.~Sannino and K.~Tuominen,
 Phys.\ Rev.\  D {\bf 71}, 051901 (2005)
 [arXiv:hep-ph/0405209].

\bibitem{Hong:2004td}
 D.~K.~Hong, S.~D.~H.~Hsu and F.~Sannino,
 Phys.\ Lett.\  B {\bf 597}, 89 (2004)
 [arXiv:hep-ph/0406200].

\bibitem{Gies:2005as}
 H.~Gies and J.~Jaeckel,
 Eur.\ Phys.\ J.\  C {\bf 46}, 433 (2006)
 [arXiv:hep-ph/0507171].

\bibitem{Braun:2005uj}
 J.~Braun and H.~Gies,
 Phys.\ Lett.\  B {\bf 645}, 53 (2007)
 [arXiv:hep-ph/0512085].

\bibitem{Braun:2006jd}
 J.~Braun and H.~Gies,
 JHEP {\bf 0606}, 024 (2006)
 [arXiv:hep-ph/0602226].

\bibitem{Dietrich:2006cm}
 D.~D.~Dietrich and F.~Sannino,
 Phys.\ Rev.\  D {\bf 75}, 085018 (2007)
 [arXiv:hep-ph/0611341].

\bibitem{Ryttov:2007sr}
 T.~A.~Ryttov and F.~Sannino,
 Phys.\ Rev.\  D {\bf 76}, 105004 (2007)
 [arXiv:0707.3166 [hep-th]].

\bibitem{Ryttov:2007cx}
 T.~A.~Ryttov and F.~Sannino,
 Phys.\ Rev.\  D {\bf 78}, 065001 (2008)
 [arXiv:0711.3745 [hep-th]].


\bibitem{Sannino:2008ha}
 F.~Sannino,
 arXiv:0804.0182 [hep-ph] 

\bibitem{Poppitz:2009tw}
  E.~Poppitz, M.~Unsal,
  JHEP {\bf 0912}, 011 (2009).
  [arXiv:0910.1245 [hep-th]].

\bibitem{Sannino:2009za}
 F.~Sannino,
  Acta Phys.\ Polon.\  {\bf B40}, 3533-3743 (2009).
  [arXiv:0911.0931 [hep-ph]].

\bibitem{Braun:2009ns}
 J.~Braun and H.~Gies,
 JHEP {\bf 1005}, 060 (2010)
 [arXiv:0912.4168 [hep-ph]].

\bibitem{Antipin:2009wr}
 O.~Antipin and K.~Tuominen,
 Phys.\ Rev.\  D {\bf 81}, 076011 (2010)
 [arXiv:0909.4879 [hep-ph]].

\bibitem{Antipin:2009dz}
  O.~Antipin, K.~Tuominen,
     [arXiv:0912.0674 [hep-ph]].

\bibitem{Jarvinen:2009fe}
 M.~Jarvinen and F.~Sannino,
 JHEP {\bf 1005}, 041 (2010)
 [arXiv:0911.2462 [hep-ph]].

\bibitem{Mojaza:2010cm}
  M.~Mojaza, C.~Pica, F.~Sannino,
  Phys.\ Rev.\  {\bf D82}, 116009 (2010).
  [arXiv:1010.4798 [hep-ph]].

 
\bibitem{Alanen:2010tg}
 J.~Alanen, K.~Kajantie and K.~Tuominen,
 Phys.\ Rev.\  D {\bf 82}, 055024 (2010)
 [arXiv:1003.5499 [hep-ph]].
 
 \bibitem{Fukano:2010yv}
 H.~S.~Fukano and F.~Sannino,
 Phys.\ Rev.\  D {\bf 82}, 035021 (2010)
 [arXiv:1005.3340 [hep-ph]].

\bibitem{Pica:2010mt}
 C.~Pica and F.~Sannino,
 arXiv:1011.3832 [hep-ph].

\bibitem{Pica:2010xq}
 C.~Pica and F.~Sannino,
 Phys.\ Rev.\  D {\bf 83}, 035013 (2011)
 [arXiv:1011.5917 [hep-ph]].

\bibitem{Frandsen:2010ej}
  M.~T.~Frandsen, T.~Pickup, M.~Teper,
  Phys.\ Lett.\  {\bf B695}, 231-237 (2011).
  [arXiv:1007.1614 [hep-ph]]

\bibitem{Ryttov:2010iz}
  T.~A.~Ryttov, R.~Shrock,
  Phys.\ Rev.\  {\bf D83}, 056011 (2011).
  [arXiv:1011.4542 [hep-ph]]

\bibitem{Chen:2010er}
  N.~Chen, T.~A.~Ryttov, R.~Shrock,
  Phys.\ Rev.\  {\bf D82}, 116006 (2010).
  [arXiv:1010.3736 [hep-ph]].


\bibitem{Ryttov:2010hs}
  T.~A.~Ryttov, R.~Shrock,
  Phys.\ Rev.\  {\bf D81}, 116003 (2010).
  [arXiv:1006.0421 [hep-ph]].

\bibitem{Ryttov:2010jt}
  T.~A.~Ryttov, R.~Shrock,
  Eur.\ Phys.\ J.\  {\bf C71}, 1523 (2011).
  [arXiv:1005.3844 [hep-ph]].

 
\bibitem{Braun:2010qs}
 J.~Braun, C.~S.~Fischer and H.~Gies,
 arXiv:1012.4279 [hep-ph].

\bibitem{Jarvinen:2010ks}
 M.~Jarvinen and F.~Sannino,
 JHEP {\bf 1102}, 081 (2011)
 [arXiv:1009.5380 [hep-ph]].

\bibitem{Sannino:2009aw}
 F.~Sannino,
 Phys.\ Rev.\  D {\bf 79}, 096007 (2009)
 [arXiv:0902.3494 [hep-ph]].

\bibitem{Ryttov:2009yw}
 T.~A.~Ryttov and F.~Sannino,
 Int.\ J.\ Mod.\ Phys.\  A {\bf 25}, 4603 (2010)
 [arXiv:0906.0307 [hep-ph]].

\bibitem{Sannino:2009qc}
 F.~Sannino,
 Phys.\ Rev.\  D {\bf 80}, 065011 (2009)
 [arXiv:0907.1364 [hep-th]].

\bibitem{Hooft} G. 't Hooft, {\it Recent Developments in Gauge Theories}, Plenum Press, 1980, 135; reprinted in {\it Unity of Forces in the Universe Vol. II}, A. Zee ed., World Scientific 1982, 1004.

\bibitem{Terning:1997xy}
 J.~Terning,
 Phys.\ Rev.\ Lett.\  {\bf 80}, 2517 (1998)
 [arXiv:hep-th/9706074].

\bibitem{Sannino:2009me}
 F.~Sannino,
 Nucl.\ Phys.\  B {\bf 830}, 179 (2010)
 [arXiv:0909.4584 [hep-th]].

\bibitem{Mojaza:2011rw}
 M.~Mojaza, M.~Nardecchia, C.~Pica and F.~Sannino,
 Phys.\ Rev.\  D {\bf 83}, 065022 (2011)
 [arXiv:1101.1522 [hep-th]].

\bibitem{Sannino:2011mr}
 F.~Sannino,
 arXiv:1102.5100 [hep-ph].

\bibitem{Sannino:2010ca}
 F.~Sannino,
 Phys.\ Rev.\  D {\bf 82}, 081701 (2010)
 [arXiv:1006.0207 [hep-lat]].

\bibitem{Sannino:2010fh}
 F.~Sannino,
 Phys.\ Rev.\ Lett.\  {\bf 105}, 232002 (2010)
 [arXiv:1007.0254 [hep-ph]].

\bibitem{DiChiara:2010xb}
 S.~Di Chiara, C.~Pica and F.~Sannino,
 arXiv:1008.1267 [hep-ph].



\bibitem{Machacek:1983tz}
 M.~E.~Machacek and M.~T.~Vaughn,
 Nucl.\ Phys.\  B {\bf 222}, 83 (1983).
 
\bibitem{Machacek:1983fi}
 M.~E.~Machacek and M.~T.~Vaughn,
 Nucl.\ Phys.\  B {\bf 236}, 221 (1984).

\bibitem{Luo:2002ti}
 M.~x.~Luo, H.~w.~Wang and Y.~Xiao,
 Phys.\ Rev.\  D {\bf 67}, 065019 (2003)
 [arXiv:hep-ph/0211440].

\bibitem{Catterall:2011pd}
 S.~Catterall, E.~Dzienkowski, J.~Giedt, A.~Joseph and R.~Wells,
 JHEP {\bf 1104}, 074 (2011)
 [arXiv:1102.1725 [hep-th]].

\bibitem{Catterall:2010nd}
 S.~Catterall,
 arXiv:1005.5346 [hep-lat].




\end{thebibliography}
\end{document}